\documentstyle[12pt]{article}

\newcommand{\resection}[1]{\setcounter{equation}{0}\section{#1}}

\newcommand{\EQ}{\begin{equation}}
\newcommand{\EN}{\end{equation}}
\newcommand{\bea}{\begin{eqnarray}}
\newcommand{\eea}{\end{eqnarray}}
\newcommand{\hs}{\hspace{0.1cm}}

\begin{document}
\setcounter{page}{0}
\topmargin 0pt
\oddsidemargin 5mm
\renewcommand{\thefootnote}{\arabic{footnote}}
\newpage
\setcounter{page}{0}
\begin{titlepage}
\begin{flushright}
ISAS/EP/94/123
\end{flushright}
\vspace{0.5cm}
\begin{center}
{\large {\bf Scattering Theory and Correlation Functions in \\
Statistical Models with a Line of Defect}}\\
\vspace{1.8cm}
{\large G. Delfino, G. Mussardo and P. Simonetti}\\
\vspace{0.5cm}
{\em International School for Advanced Studies,\\
and \\
Istituto Nazionale di Fisica Nucleare\\
34014 Trieste, Italy}\\

\end{center}
\vspace{1.2cm}

\renewcommand{\thefootnote}{\arabic{footnote}}
\setcounter{footnote}{0}

\begin{abstract}
\noindent
The scattering theory of the integrable statistical models can be generalized
to the case of systems with extended lines of defect.
This is done by adding the reflection and transmission amplitudes for
the interactions with the line of inhomegeneity to the scattering
amplitudes in the bulk. The factorization condition for the new
amplitudes gives rise to a set of Reflection-Transmission equations.
The solutions of these equations in the case of diagonal $S$-matrix in the
bulk are only those with $S =\pm 1$. The choice $S=-1$ corresponds to the
Ising model. We compute the exact expressions of the transmission and
reflection amplitudes relative to the interaction of the Majorana fermion of
the Ising model with the defect. These amplitudes present a weak-strong
duality in the coupling constant, the self-dual points being the special
values where the defect line acts as a reflecting surface. We also discuss the
bosonic case $S=1$ which presents instability properties and resonance states.
Multi-defect systems which may give rise to a band structure are also
considered. The exact expressions of correlation functions is
obtained in terms of Form Factors of the bulk theory and matrix elements of
the defect operator.
\end{abstract}

\vspace{.3cm}

\end{titlepage}

\newpage
\resection{Introduction}

In the past few years, methods and concepts of Quantum Field Theory
(QFT) have been successfully applied to the analysis of homogeneous statistical
models at and away from criticality. Besides the powerful techniques of
Conformal Field Theories which enable us to disentangle the dynamics of
the massless fluctuations of the critical points \cite{BPZ,ISZ}, the bootstrap
approach - originally developed in particle physics to describe the scattering
processes of strongly interacting particles - has proved to be one of the most
efficient way to characterize the massive excitations of the statistical
models away from criticality \cite{Zam}. Exact solutions of
massive QFT have been found along particular integrable trajectories of the
Renormalization Group, characterized by the existence of an infinite
number of conservation laws. Data easily accessible for the models associated
to the integrable trajectories are provided by their exact factorizable
$S$-matrix and mass spectrum (see, for instance [3-8]). More importantly, the
complete knowledge of the on-shell dynamics of those theories, combined
together with general properties of analyticity and relativistic invariance,
has allowed us to derive their off-shell behaviour, i.e. the computation of
correlation functions, and to make contact with the original statistical models
[9-15].

Homogeneous systems, however, are in many cases a mathematical idealization
of the real physical samples which may present instead boundary
effects and various types of inhomogeneity or defects. It is an interesting
problem in statistical mechanics to estimate the influence of the
inhomogeneities on the results obtained in pure cases and to develop the
corresponding theory. With reference to systems with boundaries, they have
been the subject of a wide investigation which has employed a large variety
of techniques [18-22]. The bootstrap approach recently developed has brought
new light on the topic and has provided quite remarkable achievements in the
understanding of QFT with boundary [23-29]. As we show in this paper,
bootstrap methods are also extremely efficient to describe integrable
statistical models with extended lines of defects. Before developing the
bootstrap theory, it is worth to briefly discuss general aspects of
statistical models with lines of defect in order to gain some insight to
their properties [28-35].

One of the main reasons for considering extended lines of inhomogeneities is
that only such kind of defects may affect the critical properties of the pure
systems. Indeed, in the opposite case where there are only a finite number of
localized inhomogeneities in the lattice, they would be eventually neutralized
by iterating the Renormalization Group transformations so that the regime
of the pure model will definitely take over.

Scaling considerations are also useful to understand in simple terms the
continuum version of the models with an infinite line of defect and to show
that they may interpolate between a bulk or a boundary statistical behaviour.
For sake of clarity, let us consider the simplest physical realization
given by a system at temperature $T$ in the bulk but heated at a
different temperature $\tilde T$ along a line placed at the $y$ axis.
This system may be equally regarded as two semi-infinite copies of the model at
temperature $T$ coupled together through the energy density at the defect
line. Its continuum properties are described by the euclidean action
\EQ
{\cal A}\,=\,{\cal A}_B \,+\,
g\,\int d^2r\,\delta(x)\,
\epsilon(r) \,\, ,
\label{action}
\EN
where ${\cal A}_B$ is the action relative to the bulk and $\epsilon(r)$ is
the energy density with scaling dimension $\nu$. The scaling dimension of the
coupling constant $g = (\tilde T - T)$ is then given by $y_g \,=\,1-\nu $.
Consequently, all those systems with an irrelevant energy operator of scaling
dimension $\nu > 1$ will exhibit the bulk critical behaviour near a defect
line. On the contrary, those models which have a relevant energy operator with
$\nu <1 $ will present a surface critical behaviour. The reason is that,
in the former case the effective coupling constant may become arbitrary
small and then the action reduces to that of the bulk theory, whereas, in the
latter case it may take arbitrary large values suppressing all the
fluctuations across the defect line between the two semi-infinite copies
which will eventually decouple.

An exception to the above pictures is given by the purely marginal case, i.e.
$\nu=1$ which is realized in the Ising model. The interesting result
obtained in the past by Bariev \cite{Bariev} and McCoy and Perk
\cite{Mc} is that the model presents a non universal critical
behaviour, with the critical indices of the magnetization operators
continuously dependent on the parameter $g$ of the action (\ref{action}).
The energy operator on the contrary remains a purely marginal operator for
all values of the coupling constant $g$ since its critical exponent $\nu$
is fixed at the Ising value of $1$ \cite{Bariev,Mc,Kad,Brown}.

Let us now turn our attention to the boostrap theory of the integrable
statistical models with extended line of defect which was originally proposed
in our previous publication \cite{DMS}. In the continuum limit and away from
criticality, such a theory can be formulated in terms of scattering processes
of the massive excitations which take place either in the bulk or on the defect
line. Hence, in addition to the bulk scattering amplitudes, we have to
consider a new set of amplitudes relative to the interaction of
the particles with the defect line. Because of the integrability condition,
they only reduce to reflection and transmission processes. It is then
convenient to associate  an extra operator $D$ to the defect line
and to formulate the dynamics in terms of algebraic relations which involve
$D$ and the operators $A_a^{\dagger}$ of the massive excitations.
The consistency of this algebra provides the unitary equations for the
transmission and reflection amplitudes while the associativity condition
gives rise to a set of cubic relations called Transmission-Reflection
Equations. Although the general solution of these equations is still lacking,
the Ising model with a line of defect can be identified as one
particular solution of them. For this model, the particles in
the bulk are given by massive Majorana fermions. The availability of an exact
resummation of the perturbative series in powers of the strength of the defect
permits to test the complementary algebraic approach referred above.
A similar approach is also used to discuss the case of free boson theory
with a line of defect, which constitutes another solution of the
Reflection-Transmission equations. The novelty of the model consists in
the presence of resonance states and instability properties by varying
the strength of the defect coupling.

As for the integrable QFT in the bulk, we will show that also for the
statistical models with a line of defect, the knowledge of the total
set of scattering amplitudes permit the full reconstruction of the theory,
i.e. the computation of the multipoint correlators. In particular,
the aforementioned non-universal critical behaviour of the Ising model can be
easily recovered by looking at the ultraviolet behaviour of the correlation
functions.

The layout of this paper is as follows. In the next section, we introduce
the defect operator $D$ and the relative algebra with the operators
of the asymptotic particles $A_a^{\dagger}$. We initially derive the unitarity
and crossing equations for the scattering amplitudes and then the
Transmission-Reflection Equations which express their factorization properties.
In Section 3, we discuss in detail one of the solutions of these equations,
i.e. the Ising model with a line of defect. In Section 4 we consider the
behaviour of the bosonic theory when coupled to a defect line. Geometrical
configurations with richer structure of inhomegeneities are considered in
Section 5. The computation of correlation functions of the model, based on
the knowledge of the Form Factors of the theory in the bulk and on
the matrix elements of the defect operator ${\cal D}$, is presented in
Section 6. Our conclusions are then summarized in Section 7.

\resection{Defect Algebra}

In the bulk, the theory of two-dimensional integrable statistical models
with a finite correlation length can be elegantly formulated in terms of
an ensemble of particle excitations in bootstrap interaction [3-8].
Although the bootstrap principle alone is in many cases sufficient to solve the
dynamics, it is also useful to rely on a more conventional approach and to
introduce an action that describes the interactions in the bulk
\EQ
{\cal A}_B\,=\,\int_{-\infty}^{+\infty} dx\,\int_{-\infty}^{+\infty} dy
\, {\cal L}_B\left(\partial_{\mu}\phi_i,\phi_i\right) \,\,\,.
\label{bulk}
\EN
In the following we assume we have completely solved the dynamics of the
corresponding theory in the Minkoswki space and as a result, we know both the
mass spectrum $\{m_a\}$ and the bulk scattering amplitudes
$S_{ab}^{cd}(\beta_{ab})$ \footnote{$\beta_{ab}=
\beta_a-\beta_b$, where $\beta_i$ is the rapidity variable of the particle
$A_i^{\dagger}$. It is related to the momenta by $p_0^{(i)}=m_i\cosh\beta_i$,
$p_1^{(i)}=m_i\sinh\beta_i$.}. The main features of the bulk
scattering processes can be briefly formulated as follows \cite{ZZ}.
To each particle of the theory we associate an operator
$A^{\dagger}_a(\beta)$ and the set of all these operators satisfy the
Faddev-Zamolodchikov algebra given by
\EQ
A_a^{\dagger}(\beta_1)\,A_b^{\dagger}(\beta_2)\,=\,S_{ab}^{cd}(\beta_{ab})\,
A_c^{\dagger}(\beta_2)\,
A_d^{\dagger}(\beta_1) \,\,\,.
\label{s}
\EN
The consistency of this algebra requires the validity of the unitarity
equations
\EQ
S_{ab}^{cd}(\beta)\,S_{cd}^{lm}(-\beta)\,=\,\delta_a^l\,\delta_b^m
\,\,\,.
\label{u}
\EN
The commutation relations (\ref{s}) should also be compatible with the
requirements of algebraic associativity expressing the factorization of
the scattering processes, i.e. they have to satisfy the Yang-Baxter equations
\EQ
S_{i_1 i_2}^{k_1 k_2}(\beta_{12})\,S_{k_1 k_3}^{j_1 j_3}(\beta_{13})\,
S_{k_2 i_3}^{j_2 k_3}(\beta_{23}) \,=\,
S_{i_1 i_3}^{k_1 k_3}(\beta_{13})\,S_{k_1 k_2}^{j_1 j_2}(\beta_{12})\,
S_{k_3 i_2}^{k_2 j_3}(\beta_{23}) \,\,\,.
\label{yb}
\EN
The analytic continuation from the $s$-channel to the $t$-channel
of the scattering process finally implies
\EQ
S_{ik}^{lj}(\beta)\,=\,S_{i\overline j}^{\overline k l}(i\pi - \beta) \,\,\,.
\label{cro}
\EN
Eqs. (\ref{s})-(\ref{cro}), eventually supported by the bootstrap equations
relative to the bound states \cite{Zam}, encode all the dynamics of the theory
in the bulk.

Let us now consider the presence of a defect line in the system, placed along
the $y$-axis. The general action of the system is given in this case by
\EQ
{\cal A}\,=\,{\cal A}_B\,+\,\int\,d^2r \, \delta(x)
{\cal L}_D\left(\phi_i,\frac{d\phi_i}{dy}\right)
\,\,\, .
\label{defect}
\EN
The new interaction, responsible for scattering processes which take place
on the defect line, will generally spoil the original integrability of
the theory: particles which hit the defect with sufficient energy
may excite internal degrees of freedom of the defect (being eventually
absorbed by it), or may give rise to production processes with multiparticle
states propagating through the two semi-infinite systems placed on the two
sides of the defect line. However, assuming that the additional interaction
along the defect line is still compatible with the existence of an
infinite number of conserved charges in involution, the dynamics drastically
simplifies and consequently is suitable for an exact analysis, as we show
in the sequel.

By translation invariance along the $y$-direction (which we here identify
with the time axis in the Minkowski space), for the theory described
by the action (\ref{defect}) we still have the conservation
of the energy but not of the momentum. Therefore we may have
scattering processes with an exchange of momentum on the defect line,
compatible though with the conservation of the energy. If in addition to
the energy other higher charges are also conserved, the scattering processes
at the defect line must be completely elastic. In particular, this means
that a particle which hits the defect line with rapidity $\beta$
can only proceed forward with the same rapidity or reverses its motion
acquiring a rapidity of $-\beta$. A further effect of the interaction
with the defect line may be a change of the label of the particle
inside its multiplet of degeneracy. The interactions of the particles
$\mid \beta;i>$ with the defect line will then be described in terms of the
transmission and reflection amplitudes, denoted respectively by
$T_{ij}(\beta)$ and $R_{ij}(\beta)$ (Fig.\,1).

The interaction of the particles at the line of inhomegeneity may be encoded
in a set of algebraic relations analogous to those which describe
the scattering processes in the bulk. In order to illustrate this,
an additional operator $D$ associated to the defect line
should be introduced in the theory\footnote{For simplicity, we discuss the
case of a defect without internal degrees of freedom and therefore carrying no
additional indices. The formulation of the more generale case is
straightforward.}. This operator may be considered in relation to an additional
particle state with zero rapidity in the entire time evolution of the system.
Its commutation relations with the creation operators $A_{i}^{\dagger}(\beta)$
associated to the asymptotic particles in the bulk are given by
\EQ
\begin{array}{lll}
A_{i}^{\dagger}(\beta)\,D & = & R_{ij}(\beta)\,
A_{j}^{\dagger}(-\beta) \,D +
T_{ij}(\beta) \,D \,A_{j}^{\dagger}(\beta) \,\,\, ;\\
D\, A_i^{\dagger}(\beta) & = & R_{ij}(-\beta)\,D\,
A_j^{\dagger}(-\beta) +
T_{ij}(-\beta) \,A_j^{\dagger}(\beta)\, D \,\,\, .
\end{array}
\label{algebra}
\EN
The first of these equations expresses the scattering of a particle that
hits the defect
coming from the semi-infinite system on the left hand side with rapidity
$\beta$. The second of (\ref{algebra}) is obtained by an analytic continuation
$\beta \rightarrow - \beta$ of the scattering amplitudes of a particle that
approaches the defect coming from the semi-infinite system on the right
hand side. The consistency condition of this algebra requires the unitarity
equations
\EQ
\begin{array}{ll}
R_{ij}(\beta)\, R_{jk}(-\beta) + T_{ij}(\beta)\,T_{jk}(-\beta) & =
\,\delta_{ik}
\,\,\, ;\\
R_{i}^{j}(\beta)\,T_{j}^{k}(-\beta) + T_{i}^{j}(\beta)\,R_{j}^{k}(-\beta) &
= \,0\,\,\,.
\end{array}
\label{unitarity}
\EN
Additional constraints emerge from the crossing relations
\EQ
\begin{array}{lll}
R_{ij}(\beta) & = & S_{\overline j l}^{k,\overline i}(2\beta)\,
R_{kl}(i\pi -\beta) \,\,\, ;\\
T_{i\bar j}(\beta) & = & T_{ij}(i\pi-\beta) \,\, .
\end{array}
\label{crossing}
\EN
The first equation in (\ref{crossing}) is obtained according to the argument
proposed in \cite{GZ,KF} which exploits the quantization of the theory in the
scheme where the time axis is placed along the $x$-axis.
With reference to the second equation in (\ref{crossing}), the transmission
channel of the process shares the same properties of ordinary scattering in
the bulk, the only difference being the occurrence of the particle $D$
with zero rapidity. Thus, it is natural to assume that in the transmission
channel the crossing symmetry is implemented in the usual way. We will assume
the validity of eqs. (\ref{crossing}) and will check that they are actually
satisfied each time we will provide explicit solutions of the scattering
theories with a line of defect.

Usually the presence of an infinite number of integrals of motion implies
not only the elasticity of all scattering processes but also their
complete factorization, i.e. an $n$-particle scattering amplitude can
be entirely expressed in terms of the elementary two-body interactions
\cite{ZZ}. A crucial step for proving the factorization property of the total
$S$-matrix is to impose the associativity condition of the algebra
(\ref{algebra}). In terms of physical process, this means that we prepare
initially an asymptotic two-particle state consisting of
$\mid A_i^{\dagger}(\beta_1)\, A_j^{\dagger}(\beta_2)>$  with
$\beta_1 > \beta_2$, and we let it scatter with the defect particle $D$
with zero rapidity. The final output of the process should be independent
from the temporal sequence of the elementary two-body interactions. Although
what we have just described looks like an ordinary three-body process of the
type that occurs in the bulk, there is however one distinguishing feature.
In fact, in the three-body processes which take place in the bulk, given
an initial state $\mid A_i^{\dagger}(\beta_1) A_j^{\dagger}(\beta_2)
A_k^{\dagger}(\beta_3)>$ identified by a set of three ordered rapidities
$\beta_1 > \beta_2 > \beta_3$, there is an {\em unique} final state given by
the reverse ordering of the rapidities and possible exchange of the internal
indices among the particles. On the contrary, for the scattering processes on
the defect line we may have four possible final states, namely: ({\em a})
the state with both particles reflected by the defect line; ({\em b})
the state with both particles transmitted; ({\em c} and {\em d}) the states
with one particle reflected whereas the other one transmitted.
The final states may also differ from the initial one for the exchange of the
internal indices and the above four possibilities give rise to a
set of Reflection-Transmission (RT) equations shown in Fig.\,2.

The first of these (Fig.\,2.a) coincides with the well-known boundary
equations already analysed in \cite{Cherednik,Sklyanin,GZ},
\EQ
S_{ac}^{ef}(\beta_1-\beta_2) R_{fg}(\beta_2) S_{ge}^{dh}(\beta_1+\beta_2)
R_{gb}(\beta_1)\,=\,R_{ah}(\beta_2) S_{ch}^{fe}(\beta_2+\beta_1)
R_{fg}(\beta_2) S_{eg}^{bd}(\beta_1-\beta_2)\,\,\, .
\label{refl1}
\EN
The RT equations associated with the configurations of Figs. (2.b), (2.c) and
(2.d) are given respectively by
\EQ
\begin{array}{l}
S_{ac}^{lm}(\beta_1-\beta_2) \,T_{lb}(\beta_1) \,T_{md}(\beta_2)\,=\,
S_{ml}^{bd}(\beta_1-\beta_2) \,T_{cm}(\beta_2) \,T_{dl}(\beta_1)
\,\,\,;\\
S_{ac}^{fe}(\beta_1-\beta_2) \,T_{fb}(\beta_1)\,R_{ed}(\beta_2) \,=\,
R_{ce}(\beta_2)\,S_{ae}^{fd}(\beta_1+\beta_2)\,T_{fb}(\beta_1)\,\,\,;\\
S_{ac}^{fe}(\beta_1-\beta_2) \,R_{fg}(\beta_2) \,S_{ge}^{dh}(\beta_1+\beta_2)
\,T_{hb}(\beta_1)\,=\,T_{ab}(\beta_1) \,R_{cd}(\beta_2) \,\,\, . \\
\end{array}
\label{YB}
\EN
Although a general solution of these equations is still lacking, it is
easy to see that they become extremely restrictive once applied to
QFT with a non-degenerate spectrum, i.e. those which have a diagonal $S$-matrix
in the bulk. In fact, whereas eq.\,(\ref{refl1}) and the first in (\ref{YB})
are identically satisfied, the last two equations in (\ref{YB}) become in
this case
\EQ
\begin{array}{l}
S_{ab}(\beta_a\,+\beta_b) \,=\, S_{ab}(\beta_b-\beta_a) \,\, ,\\
S_{ab}(\beta_a\,+\,\beta_b)\,S_{ab}(\beta_a\,-\,\beta_b)\,=\,1 \,\, ,
\end{array}
\label{cr}
\EN
Hence, from the first equation in (\ref{cr}) we see that the $S$-matrix in the
bulk has to be a constant and from the second equation (or equivalently from
the unitarity condition) this constant is fixed to be $\pm 1$. Thus we conclude
that the only integrable QFT with diagonal $S$-matrix in the bulk and
factorizable scattering in the presence of the defect line are those
associated to generalized-free theories.

Obviously this restriction on the bulk $S$-matrix does not apply when one
considers purely reflective theories because they are ruled only by
equations (\ref{refl1}). Non-trivial solutions of these equations have been
analysed for several models and they provide explicit examples of QFT with
boundary [23-29], some of them of relevant importance in statistical mechanics.

\resection{Ising Model with a Line of Defect}

As we have seen at the end of the previous section, the validity of the
Transmission-Reflection Equations in the case of non-degenerate mass spectrum
selects $S = -1$ as a possible scattering matrix in the bulk. This solution
can be identified as the scattering amplitude of the particle excitations of
the Ising model, given by the massive Majorana fermions \cite{IsingM}. The
Lagrangian density of the continuum theory in the bulk is given by
\EQ
{\cal L}_B\,=\,\overline\Psi(x,t)\left(i\gamma^{\mu}\partial_{\mu} - m\right)
\Psi(x,t) \,\,\, .
\label{Majorana}
\EN
In the Majorana representation, given by $\gamma^0 = \sigma_2$,
$\gamma^1 = -i \sigma_1$, the fermionic field $\Psi(x,t)$ is real, i.e.
$\Psi^{\dagger}(x,t) = \Psi(x,t)$. The physical content
of the model, as defined by the Lagrangian (\ref{Majorana}), does not
depend on the sign in front of the mass term since it can be altered by
the transformation $\Psi \rightarrow \Psi$; $\overline\Psi \rightarrow
-\overline\Psi$ of the fermionic field. As it is well known, the mass $m$ is a
linear measurement of the deviation of the temperature with respect to
the critical one
\EQ
m = 2\pi (T - T_c) \,\, ,
\label{mass}
\EN
and the symmetry $m \rightarrow -m$ simply expresses the self-duality of
the model. In the high temperature phase given by $m > 0$, the vacuum
expectation value of the magnetization operator $\sigma$ vanishes,
whereas, the corresponding quantity of the disorder operator $\mu$ is different
from zero. Under the duality transformation, the role of order and disorder
operators is reversed whereas for the energy operator $\epsilon$, given
by $\epsilon = i  \overline\Psi \Psi$, we simply have a change in its sign.

On a square lattice, the Ising model with a line of defect can be realized
in two different ways (Fig.\,3). The first is the chain geometry with bulk
coupling constants $J$ and modified couplings $\tilde J$ parallel to the
defect line. The second one is the ladder geometry, with the modified
set of couplings placed in the perpendicular direction. Since the two geometric
realizations are related by Kramer-Wannier duality symmetry, from now on we
can restrict our attention to one of them, say the chain geometry.
In the continuum formulation, the defect line introduces the additional
term\footnote{The exact relationship between $g$ and the lattice
coupling constants will be established in section 6 by comparing correlation
functions computed in the lattice and in the continuum formulation.}
\EQ
{\cal L}_D\,=\, -g\,\delta(x) \overline\Psi(x,t)\Psi(x,t) \,\,
\label{defising}
\EN
to the Lagrangian (\ref{Majorana}). The new interaction is purely marginal and
therefore the beta-function associated to the coupling constant $g$ is
identically zero. The marginality of the interaction has as a consequence
that the theory presents a non-universal ultraviolet behaviour in the
magnetization sector, with the critical exponent of the magnetization
operators which depend continuously on the parameter $g$, whereas the energy
operator always keeps its original value of $1$ of the Ising model
\cite{Bariev,Mc,Kad,Brown}.

In this section we are interested in determining the reflection $R(\beta)$
and transmission $T(\beta)$ amplitudes for the scattering of the fermion with
the defect line, i.e. the $S$-matrix elements between initial and final states
$u(p_i)$ and $\overline u(p_f)$ with $p_f = \pm p_i$. To this aim,
let us consider the perturbative series of the Green function
of the fermion field $\Psi$ based on the Feynman rules
\vspace{2mm}

\begin{picture}(210,25)
\put(50,8){\vector(1,0){25}}
\put(54,0){\makebox(0,0){$p$}}
\put(75,8){\line(1,0){15}}
\put(86,1){\makebox(0,0){$p'$}}
\put(100,8){\makebox(0,0)[l]{$= i (2\pi)^2\, \delta^2 (p - p') \,
\frac{\not p + m}{p^2 - m^2 + i\epsilon}$}}
\end{picture}
\begin{picture}(250,25)
\put(50,8){\vector(1,0){11}}
\put(61,8){\line(1,0){10}}
\put(75,8){\circle*{8}}
\put(79,8){\vector(1,0){11}}
\put(90,8){\line(1,0){10}}
\put(105,8){\makebox(0,0)[l]{$ = -\,i g \,2\pi\,\delta (p_0 - p_0')$}}
\end{picture}

\vspace{1mm}
\noindent
For the self-energy entering the exact propagator we have
the following series of diagrams
\vspace{3mm}

\begin{picture}(420,50)
\put(28,25){\vector(1,0){11}}
\put(39,25){\line(1,0){10}}
\put(69,25){\oval(40,25)}
\put(69,25){\makebox(0,0){$\Sigma$}}
\put(89,25){\vector(1,0){11}}
\put(100,25){\line(1,0){10}}
\put(118,25){\makebox(0,0){$=$}}
\put(129,25){\vector(1,0){8}}
\put(137,25){\line(1,0){3}}
\put(143,25){\circle*{6}}
\put(146,25){\vector(1,0){8}}
\put(154,25){\line(1,0){3}}
\put(168,25){\makebox(0,0){+}}
\put(180,25){\vector(1,0){8}}
\put(188,25){\line(1,0){3}}
\put(194,25){\circle*{6}}
\put(197,25){\vector(1,0){8}}
\put(205,25){\line(1,0){3}}
\put(211,25){\circle*{6}}
\put(214,25){\vector(1,0){8}}
\put(222,25){\line(1,0){3}}
\put(229,25){\makebox(0,0)[l]{+}}
\put(241,25){\vector(1,0){8}}
\put(249,25){\line(1,0){3}}
\put(255,25){\circle*{6}}
\put(258,25){\vector(1,0){8}}
\put(266,25){\line(1,0){3}}
\put(272,25){\circle*{6}}
\put(275,25){\vector(1,0){8}}
\put(283,25){\line(1,0){3}}
\put(289,25){\circle*{6}}
\put(292,25){\vector(1,0){8}}
\put(300,25){\line(1,0){3}}
\put(312,25){\makebox(0,0)[l]{$+ \cdots $}}
\end{picture}

\noindent
where we have to integrate on the spatial component of the momentum
running in the internal lines. The integral on the intermediate state is
given by

\vspace{1mm}
\begin{picture}(250,50)
\put(34,25){\circle*{6}}
\put(37,25){\vector(1,0){8}}
\put(43,15){\makebox(0,0){$k$}}
\put(45,25){\line(1,0){3}}
\put(51,25){\circle*{6}}
\put(64,25){\makebox(0,0){=}}
\put(76,25){\makebox(0,0)[l]
{$ (-ig)^2\, i\,\delta(k_0 - p_0)\,\int\frac{dk^1}{2\pi}\,
\frac{\not k +m}{k^2 - m^2 + i\epsilon} \,
= \,-g^2\, \delta(k_0 - p_0)\,\frac{p_0 \gamma^0 + m}{2 \omega}
\hspace{3mm}.$}}
\end{picture}

\noindent
In the above quantity we have discarded by parity the infinity related to the
linear term in $k$. With this prescription, the geometric series for
$\Sigma$ is finite and can be expressed in a closed form as
\EQ
\Sigma(p_0)\,=\,2\pi\,i\, \delta(p_0-p_0')\,
\sin\chi \,\frac{\omega - i \frac{g}{2} (p_0 \gamma^0 - m)}
{\omega - i m\, \sin\chi} \,\, ,
\EN
where
\[
\omega\,=\,\sqrt{p_0^2 - m^2} \hspace{3mm},\hspace{5mm}
\sin\chi\,=\, -\,\frac{g}{1 + \frac{g^2}{4}} \,\,\, .
\]
We can now apply the usual LSZ reduction formulae, and for the transmission
and reflection amplitudes defined by
\[
{}_{out}<\beta'\mid\beta>_{in}\,=\,2\pi \delta(\beta - \beta')\,T(\beta,g) +
2\pi \delta(\beta + \beta')\,R(\beta,g) \,\, ,
\]
we have
\begin{eqnarray}
T(\beta,g) & = & \frac{\cos\chi\,\sinh\beta}{\sinh\beta - i \sin\chi}
\,\, ,
\nonumber\\
&&
\label{rt}\\
R(\beta,g) & = & i \, \frac{\sin\chi\,\cosh\beta}
{\sinh\beta - i \sin\chi} \,\,\, .\nonumber
\end{eqnarray}
The transmission amplitude also contains the disconnected part relative to
the free motion.

Before commenting on the properties of these amplitudes, it is interesting
to present an alternative derivation of (\ref{rt}). This is obtained by
implementing the algebra (\ref{algebra}) on the creation operators of the
fermion field. Let $\Psi_{\pm}(x,t)$ be the solutions of the free Dirac
equation in the two intervals $x>0$ and $x<0$,
i.e.
\EQ
\Psi(x,t)\,=\,\theta(x)\,\Psi_+(x,t) \,+\,\theta(-x)\,\Psi_-(x,t)
\,\, ,
\label{full}
\EN
with the value at the origin given by $
\Psi(0,t)\,=\,\frac{1}{2}\left(\Psi_+(0,t)\,+\,\Psi_-(0,t)\right)
$. The mode expansion of the two components of the fields $\Psi_{\pm}(x,t)$
is expressed as
\begin{eqnarray}
\psi_{(\pm)}^{(1)}(x,t) & = &
\int
\frac{d\beta}{2\pi} \left[
\omega e^{\frac{\beta}{2}}\, A_{(\pm)}(\beta)\, e^{-i m(t \cosh\beta
- x\sinh\beta)} \,+\,
\overline\omega e^{\frac{\beta}{2}}\, A_{(\pm)}^{\dagger}(\beta)\,
e^{i m(t \cosh\beta - x\sinh\beta)} \right]
\label{mode}\\
\psi_{(\pm)}^{(2)}(x,t) & = & -
\int
\frac{d\beta}{2\pi} \left[
\overline\omega e^{-\frac{\beta}{2}}\, A_{(\pm)}(\beta)\, e^{-i m
(t \cosh\beta - x\sinh\beta)} \,+\,
\omega e^{-\frac{\beta}{2}}\, A_{(\pm)}^{\dagger}(\beta) \,
e^{i m (t\cosh\beta - x\sinh\beta)}\right] \,\, ,\nonumber
\end{eqnarray}
with $\omega=\exp(i\pi/4)$, $\overline\omega=\exp(-i\pi/4)$.
The operators $A_{\pm}(\beta)$ and $A^{\dagger}_{\pm}(\beta)$ satisfy the
usual anti-commutation relations of a free fermion
\EQ
\left\{A_{\pm}(\beta),A_{\pm}^{\dagger}(\theta)\right\}\,=\,2\pi\,
\delta(\beta-\theta) \,\, ,
\EN
although they are not all independent. They are related to each other
by the conditions at $x=0$ which arise from applying the eqs. of motion to
(\ref{full}), i.e.
\EQ
\begin{array}{lll}
(\psi_{+}^{(2)} - \psi_{-}^{(2)})(0,t) & = & \frac{g}{2} (\psi_{+}^{(1)} +
\psi_{-}^{(1)})(0,t) \,\,\, ;\\
(\psi_{+}^{(1)} - \psi_{-}^{(1)})(0,t) & = & \frac{g}{2} (\psi_{+}^{(2)} +
\psi_{-}^{(2)} )(0,t)\,\,\,.
\end{array}
\label{bc}
\EN
These equations are equivalent to the relationship between the modes
\EQ
M\,\,
\left(\begin{array}{c}
A_-^{\dagger}(\beta) \\
A_+^{\dagger}(-\beta)
\end{array}
\right)
\,=\,
N\,\,
\left(\begin{array}{c}
A_-^{\dagger}(-\beta) \\
A_+^{\dagger}(\beta)
\end{array}
\right) \,\, ,
\EN
where
\[
M\,=\,\left(
\begin{array}{ll}
\omega e^{-\frac{\beta}{2}} + \frac{g}{2} \overline\omega e^{\frac{\beta}{2}} &
\,\,\,
-\omega e^{\frac{\beta}{2}} + \frac{g}{2} \overline\omega e^{-\frac{\beta}{2}}
\\
\omega e^{-\frac{\beta}{2}} + \frac{2}{g} \overline\omega e^{\frac{\beta}{2}} &
\,\,\,
\omega e^{\frac{\beta}{2}} - \frac{2}{g} \overline\omega e^{-\frac{\beta}{2}}
\end{array}
\right)
\hspace{3mm} ;
\]
\[
N\,=\,
\left(
\begin{array}{ll}
-\omega e^{\frac{\beta}{2}} - \frac{g}{2} \overline\omega e^{-\frac{\beta}{2}}
&
\omega e^{-\frac{\beta}{2}} - \frac{g}{2} \overline\omega e^{\frac{\beta}{2}}
\\
-\omega e^{\frac{\beta}{2}} - \frac{2}{g} \overline\omega e^{-\frac{\beta}{2}}
&
-\omega e^{-\frac{\beta}{2}} + \frac{2}{g} \overline\omega e^{\frac{\beta}{2}}
\end{array}
\right)
\,\,\, .
\]
Hence,
\EQ
\left(\begin{array}{c}
A_-^{\dagger}(\beta) \\
A_+^{\dagger}(-\beta)
\end{array}
\right)
\,=\,
M^{-1}\,N\,\,
\left(\begin{array}{c}
A_-^{\dagger}(-\beta) \\
A_+^{\dagger}(\beta)
\end{array}
\right)
\,=\,
\left(\begin{array}{ll}
R(\beta,g) & T(\beta,g) \\
T(\beta,g) & R(\beta,g)
\end{array}
\right)\,
\left(\begin{array}{c}
A_-^{\dagger}(-\beta) \\
A_+^{\dagger}(\beta)
\end{array}
\right)
\EN
with $R(\beta,g)$ and $T(\beta,g)$ given in (\ref{rt}). Note that, although
the boundary conditions (\ref{bc}) are both linear in $g$, there is however a
feedback between the two components of the fermionic field. The final
dependence from the coupling constant is then expressed in terms of
trigonometric functions of the auxiliary angle $\chi$.

Given the explicit expressions of the amplitudes (\ref{rt}), it is easy to
check that they satisfy the unitarity and crossing equations (\ref{unitarity})
and (\ref{crossing}). They present several interesting features. Firstly,
by taking their sum and difference we obtain
\[
e^{2i\delta_0}  \equiv T(\beta,g) + R(\beta,g) \,=\,
\frac{\sinh\frac{1}{2}\left(\beta + i \chi\right)}
{\sinh\frac{1}{2}\left(\beta - i \chi\right)} \,\,\, ;
\]
\[
e^{2i\delta_1} \equiv T(\beta,g) - R(\beta,g) \,=\,
\frac{\cosh\frac{1}{2}\left(\beta - i \chi\right)}
{\cosh\frac{1}{2}\left(\beta + i \chi\right)}\,\, ,
\]
which can be considered as partial-wave phase shifts, with
$\delta_0$ and $\delta_1$ crossed functions of each other. Secondly, notice
that as functions of the coupling constant $g$, they satisfy a strong-weak
duality given by
\EQ
T\left(\beta,\frac{4}{g}\right)\,= \,-\,T(\beta,g)
\hspace{3mm} ,
\hspace{6mm}
R\left(\beta,\frac{4}{g}\right)\,=\,R(\beta,g) \,\,\, .
\EN
At the self-dual points $g = \pm 2$ the transmission amplitude vanishes and
therefore the defect line behaves as a pure reflecting surface. From
the unitarity equations (\ref{unitarity}), the corresponding reflection
amplitudes $R(\beta,\pm 2)$ become pure phases, as can be explicitly seen
by their equivalent expressions
\EQ
R(\beta,\pm 2)\,=\,-\frac{\cosh\frac{\beta}{2} \pm i \sinh\frac{\beta}{2}}
{\cosh\frac{\beta}{2} \mp i \sinh\frac{\beta}{2}}\,\,\,.
\EN
They coincide with the reflection amplitudes of the Ising model with fixed
and free boundary conditions respectively, as determined in \cite{GZ}.
To establish directly the pure reflecting properties of the defect line
at the self-dual points, let us analyse more closely the decoupling which
occurs in the boundary conditions when $g=\pm 2$. For $g = 2$,
the boundary conditions (\ref{bc}) become
\EQ
\begin{array}{lll}
(\psi_{+}^{(2)} - \psi_{-}^{(2)})(0,t) & = & (\psi_{+}^{(1)} +
\psi_{-}^{(1)})(0,t) \,\,\, ;\\
(\psi_{+}^{(1)} - \psi_{-}^{(1)})(0,t) & = & (\psi_{+}^{(2)} +
\psi_{-}^{(2)} )(0,t)\,\, ,
\end{array}
\label{bc2}
\EN
and taking their sum and difference, they can be written as
\EQ
\begin{array}{lll}
(\psi_{+}^{(2)} - \psi_{+}^{(1)})(0,t) & = & 0
\,\,\,;\\
(\psi_{-}^{(2)} + \psi_{-}^{(1)})(0,t) & = & 0 \,\,\,.
\end{array}
\label{bc22}
\EN
For $g=-2$, the original boundary conditions (\ref{bc}) are reduced instead to
\EQ
\begin{array}{lll}
(\psi_{+}^{(2)} + \psi_{+}^{(1)})(0,t) & = & 0 \,\,\, ;\\
(\psi_{-}^{(2)} - \psi_{-}^{(1)})(0,t) & = & 0 \,\,\,.
\end{array}
\label{bc-2}
\EN
Equations (\ref{bc22}) and (\ref{bc-2}) explicitly show that the two
semi-infinite systems across the defect line are completely decoupled, and
each of them can be treated as a QFT in the presence of pure reflecting
surface whose role is to supply the appropriate boundary conditions
\cite{GZ}. At first sight, though, one may be surprised by the asymmetric
form assumed by the eqs. (\ref{bc22}) and (\ref{bc-2}) which treat
differently the two fermionic fields $\Psi_{\pm}(x,t)$. However, this
asymmetry has a physical origin. By means of the mode expansion (\ref{mode}),
the first equation in (\ref{bc22}) and (\ref{bc-2}) can be used to determine
directly the reflection amplitudes $R(\beta,\pm 2)$. By the same token, using
the second equation in (\ref{bc22}) and (\ref{bc-2}) we find $R(-\beta,\pm 2)$,
instead. But, this is physically correct, the reason being that, in order to
have a reflection of a particle described by $\Psi_+(x,t)$ with the defect
(boundary) line, this particle must approach the origin with positive
rapidity $\beta$. On the contrary, a reflection of a particle described
by $\Psi_-(x,t)$ with the defect (boundary) line is only realized for
negative values of its rapidity.

Further support of the identification of $R(\beta,\pm 2)$ with the
reflection amplitudes of the Ising model with fixed and free boundary
conditions comes from the analysis of the relationship between the lattice
and the continuum formulations of the chain geometry, which will be
established in Section 6. Anticipating the result, this is provided by
the formula
\EQ
\sin\chi\,=\,\tanh 2(J-\tilde J) \,\,\,.
\EN
Hence, the condition $\sin\chi =-1$ corresponds to a coupling constant
$\tilde J$ along the defect line infinitely larger (and positive) than the
coupling constant $J$ of the bulk. As a consequence, the spins along the
defect line are frozen into a fixed boundary condition. On the other hand,
the condition $\sin \chi = 1$ is obtained in the anti-ferromagnetic limit
$\tilde J \rightarrow -\infty$ where the spins along the defect line are
aligned in antiparallel configurations. Since the nearby spins couple to
a surface with vanishing magnetization, this situation corresponds to
the free boundary conditions \cite{def2}.

Let us now turn our attention to the analytic structure of the reflection
and transmission amplitudes. For negative values of $g$, the interaction with
the defect line is attractive and consequently the theory presents a bound
state with binding energy $e_b = m\cos\chi$. It is quite instructive to
calculate the transmission and reflection amplitudes $T_b(\beta)$ and
$R_b(\beta)$ relative to the scattering of the fermion with the excited state
present on the defect line. The first thing to observe is that both amplitudes
$R(\beta)$ and $T(\beta)$ present a pole singularity at
$\beta = i \chi$ and $\beta = i (\pi - \chi)$. The reflection amplitude
$R(\beta)$ has positive residue at both locations, given by $i \sin \chi$. On
the other hand, $T(\beta)$ presents a positive residue with the same value as
$R(\beta)$ at $\beta = i \chi$ and a negative residue $- i \sin \chi$ at the
other pole $\beta = i (\pi - \chi)$. The problem of identifying which one of
the two poles corresponds to the bound state is solved by selecting the
singularity with positive residue in both amplitudes. This is the pole at
$\beta = i \chi$. The relative binding energy is positive, as it should be.
To recover the transmission and reflection amplitudes relative to the excited
state, we have to impose the commutativity of the graphs shown in Fig.\,4.
Since the $S$-matrix in the bulk is $-1$, the reflection amplitude
$R_b(\beta)$ coincides with the original one i.e. $R_b(\beta) = R(\beta)$
whereas the transmission amplitude is given by $T_b(\beta) = - T(\beta)$.
If we again identify the singularity associated to a bound state as that pole
with a positive residue in both channels, we see that for the defect bound
state amplitudes the role of the two poles has been reversed! Namely, the pole
which corresponds to the bound state in $R_b(\beta)$ and $T_b(\beta)$ is now
located at $\beta = i (\pi - \chi)$ and is relative to the original ground
state of the defect line\footnote{Note that the presence of the transmission
amplitude has been quite crucial in order to discriminate which one of the two
poles with positive residue in the reflection channel corresponds to the bound
state. In the pure reflecting situation, as for instance may be the case of the
Ising model with a boundary magnetic field considered in \cite{GZ}, the
occurrence of positive residue at both poles in the reflection amplitude
and a misinterpretation of their role could in fact lead to a paradoxical
hierarchy of bound states obtained by applying iteratively the boundary
bootstrap equations.}.

As a final remark of this section, the marginal nature of the defect
interaction in the Ising model can be also inferred by looking at the
high-energy limit of the amplitudes. For large values of $\beta$ we have
\EQ
T(\beta) \sim \cos\chi \,\,\, , \hspace{5mm}
R(\beta) \sim i \sin\chi \,\,\,.
\EN
Hence, except for the special values of the coupling constant $g$ where one
of the two quantities vanish, both amplitudes are always simultaneously
present. Since this limit probes the short distance scales of the model, we
see that its critical properties of bulk and surface behaviour are
simultaneously present\footnote{For an irrelevant interaction which leads to
a bulk critical behaviour near the defect line, we expect in fact a vanishing
of the reflection amplitude in the high-energy limit. On the contrary, for a
relevant interaction, the system should show a purely surface critical
behaviour characterized by the vanishing of the transmission amplitude.}.

\resection{Bosonic Theory with a Line of Defect}

Another solution of the Reflection-Transmission equations is provided by the
massive free bosonic theory with the $S$-matrix in the bulk given by $S=1$.
As an example of a bosonic theory with a line of defect, we consider
the model described by the Lagrangian
\EQ
{\cal L} \,=\, \frac{1}{2} \left[
(\partial_{\mu}\varphi)^2 - m^2 \varphi^2 - g\, \delta(x) \varphi^2\right]
\,\,\,.
\label{lab}
\EN
For the equation of motion we have
\EQ
\left[\Box + m^2 + g\, \delta(x) \right] \varphi = 0 \,\,\, .
\EN
As for the Ising model, one can obtain the reflection and transmission
amplitudes by an exact resummation of the perturbative series in the
coupling constant $g$. The calculations are analogous to the fermionic case,
and rather than repeating them here, we prefer to exploit the algebraic
approach directly. The solution of the equation of motion may be written
as
\EQ
\varphi(x,t) = \theta(x) \varphi_+(x,t) + \theta(-x) \varphi_-(x,t)\,\, ,
\label{gue}
\EN
where the mode decomposition of the two fields $\varphi_{\pm}(x,t)$ is given
by
\EQ
\varphi_{(\pm)}(x,t) =
\int
\frac{d\beta}{2\pi} \left[
A_{(\pm)}(\beta)\, e^{-i m(t \cosh\beta
- x\sinh\beta)} \,+\,
A_{(\pm)}^{\dagger}(\beta)\,
e^{i m(t \cosh\beta - x\sinh\beta)} \right]\,\,\,.
\label{modephi}
\EN
The operators $A_{\pm}(\beta)$ and $A^{\dagger}_{\pm}(\beta)$ satisfy the
usual commutation relations of a free massive boson
\EQ
\left[A_{\pm}(\beta),A_{\pm}^{\dagger}(\theta)\right]\,=\,2\pi\,
\delta(\beta-\theta) \,\,\,.
\EN
The interaction along the defect however makes them not linearly independent.
In fact, substituting eq.\,(\ref{gue}) into the equation of motion, the
latter is equivalent to the boundary conditions
\begin{eqnarray}
\varphi_+(0,t) - \varphi_-(0,t) & = & 0 \,\,\,;\\
\frac{\partial}{\partial x} (
\varphi_+(0,t) - \varphi_-(0,t) ) & = &\frac{g}{4 m}
(\varphi_+(0,t) + \varphi_-(0,t)) \,\, ,\nonumber
\end{eqnarray}
which, in terms of the mode, can be written as
\EQ
\left(\begin{array}{c}
A_-^{\dagger}(\beta) \\
A_+^{\dagger}(-\beta)
\end{array}
\right)
\,=\,
\left(\begin{array}{ll}
R(\beta,g) & T(\beta,g) \\
T(\beta,g) & R(\beta,g)
\end{array}
\right)\,
\left(\begin{array}{c}
A_-^{\dagger}(-\beta) \\
A_+^{\dagger}(\beta)
\end{array}
\right) \,\,\, .
\EN
The transmission and reflection amplitudes in the above formula are given by
\begin{eqnarray}
T(\beta,g) & = & \frac{\sinh\beta}{\sinh\beta + i g/4 m}
\,\, ,
\nonumber\\
&&
\label{rtb}\\
R(\beta,g) & = & - \, \frac{i g/4 m}
{\sinh\beta + i g/4 m} \,\,\, .\nonumber
\end{eqnarray}
These amplitudes satisfy the unitarity and crossing equations (\ref{unitarity})
and (\ref{crossing}). It is easy to see that by substituting $\sinh\beta$
in (\ref{rtb}) with the linear momentum $k$, the two resulting amplitudes are
the same as those obtained in one-dimensional quantum mechanics for the
scattering in a $\delta$-function potential (see, for instance, \cite{QM}).
However, due to the relativistic nature of the QFT, there is an important
difference between the two cases, as shown by the analysis which follows on
the pole structure of the amplitudes (\ref{rtb}).

For the $2\pi i$ periodicity of the amplitudes, we can restrict our attention
to the strip $-i\pi \leq \beta \leq i\pi$. Let us consider initially the
case when $g$ is a positive quantity. As long as $g$ satisfies the condition
$ 0 < g < 4 m$, there are two poles on the negative imaginary axis relative
to the unphysical sheet. By increasing the value of $g$ they approach each
other, and there is a critical value $g_{c_1} = 4m$ where they collide at
position $\beta = - i\pi/2$. Additional increment of the coupling constant
causes the poles to move in the complex strip keeping their imaginary part
equal to $-i\pi/2$ but acquiring a real part (Fig.\,5). In terms of QFT,
this means that the bosonic theory with a coupling constant of the defect
line larger than $4 m$ presents two resonance states. As $g$ grows, these
poles move to infinity, and in the limit $g \rightarrow \infty$, the defect
line acts as a pure reflecting surface. Indeed, the transmission amplitude
vanishes, whereas the reflection amplitude expresses the fixed boundary
condition $\varphi(0,t) = 0$.

Let us now analyse the case when $g$ is a negative quantity. In the range
$ -4m < g < 0$, the amplitudes present two poles placed on the positive
imaginary axis relative to the physical sheet. The closest one to the
origin can be interpreted as a defect bound state. By decreasing $g$, these
two poles approach each other until they finally collide at $\beta = i \pi/2$
for the critical value $g_{c_2} = - 4 m$. Further decrement of the
coupling constant makes them move in the complex strip with an
imaginary part equals to $i \pi/2$ and with a real component which increases
by decreasing g. However, these poles are now located in the physical
strip and therefore the theory presents instability properties. The
easiest way to explicitly illustrate this instability is to consider the
analytic continuation $\beta \rightarrow \left(i\frac{\pi}{2} - \beta\right)$
in $R(\beta)$. As discussed in Section 6, the resulting quantity
$\hat R(\beta)$, given by
\EQ
\hat R(\beta,g)\, = \, - \, \frac{ g/4 m}
{\cosh\beta + g/4 m} \,\, ,
\label{rtbe}
\EN
can be interpreted as the amplitude relative to the emission of a pair of
particles with momentum $\beta$ and $-\beta$ from the defect line placed
along the $x$-axis \cite{GZ}. Then, for $g < -4m$,
$\hat R(\beta)$ presents a pole for real values of $\beta$ that induces a
spontaneous emission of pairs of particles. The occurrence of such processes
obviously spoils the stability of the theory.

In light of the above results, we can summarize the discussion by saying
that the QFT associated to the Lagrangian (\ref{lab}) makes sense only for
values of $g$ in the range $ - 4m < g \leq \infty$. In a path integral
approach to the problem, it is easy to see that there
may be a competition in the Lagrangian (\ref{lab}) between the genuine mass
term and the defect interaction. Adopting the interpretation of the
$\delta$-function interaction as a suitable limit of a constant potential in
the strip $(-\epsilon,\epsilon)$ around the origin, when $g$ is sufficiently
positive in this interval, we may have an effective mass of the field $\varphi$
in this strip higher than the threshold mass $m$ in the bulk. This
produces the resonance poles in the transmission and reflection amplitudes.
Viceversa, for negative values of $g$, the effective mass of the field
$\varphi$ in the tiny interval around the origin is smaller than the mass
gap in the bulk and it decreases until it vanishes at $g = -4m$. After this
value it becomes imaginary, giving rise to the instability property previously
discussed.

It is likewise interesting to understand the different behaviour of the
bosonic and the fermionic theories in terms of the coupling constant.
The reason is that the physical content of the fermionic model does not
depend on the sign of the mass term, which enters linearly
in the action. Therefore, by varying the coupling constant $g$, there is
no a real competition with the genuine mass term in the action, so that
the fermionic model cannot present instabilities or resonance states.
In fact, crossing the critical values $g=\pm 2$, the poles simply interchange
their positions, i.e. the weak coupling regime swaps with the strong coupling
one.

As a last comment on the bosonic theory analysed in this section, the defect
interaction is associated to an irrelevant operator and therefore the defect
line should be completely transparent in the ultraviolet limit. Indeed,
taking the the high-energy limit $\beta \rightarrow \infty$ of the amplitudes
(\ref{rtb}), the reflection amplitude vanishes whereas the transmission
amplitude is identically equal to $1$.

\resection{Models with Multi-Defect Lines}

The solutions so far determined for the fermionic and bosonic theories
in the presence of a single line of defect can be generalized and
geometrical situations with a richer structure of defect lines can be also
included. In this section, we analyse the case of two parallel lines of defect,
and then the quantization conditions induced by a periodic array of defects.
Due to the different behaviour of the fermionic and bosonic theories, it is
convenient to discuss them separately.

\subsection{Fermionic Theory}

Let us initially consider the Ising model with two parallel lines of defect,
one placed at the origin along the $y$-axis with strength $g_1$, the other
shifted by $a$ and with strength $g_2$. In the fermionic formulation of the
model, the field $\Psi(x,t)$ has a free motion in each of the three intervals
$I_- \equiv(-\infty,0)$, $I_0 \equiv (0,a)$ and $I_+ \equiv (a,+\infty)$
separated by the two defect lines. Therefore in each of the three intervals the
field $\Psi(x,t)$ admits the usual decomposition in modes and the role of the
defect lines is to provide the boundary conditions at the edges of the
intervals.
The first of them is at $x=0$ and is given by
\EQ
\begin{array}{lll}
(\psi_{0}^{(2)} - \psi_{-}^{(2)})(0,t) & = & \frac{g_1}{2} (\psi_{0}^{(1)} +
\psi_{-}^{(1)})(0,t) \,\,\, ;\\
(\psi_{0}^{(1)} - \psi_{-}^{(1)})(0,t) & = & \frac{g_1}{2} (\psi_{0}^{(2)} +
\psi_{-}^{(2)} )(0,t)\,\, ,
\end{array}
\label{bc1}
\EN
whereas for the second boundary condition at $x=a$ we have
\EQ
\begin{array}{lll}
(\psi_{+}^{(2)} - \psi_{0}^{(2)})(a,t) & = & \frac{g_2}{2} (\psi_{+}^{(1)} +
\psi_{0}^{(1)})(a,t) \,\,\, ;\\
(\psi_{+}^{(1)} - \psi_{0}^{(1)})(a,t) & = & \frac{g_2}{2} (\psi_{+}^{(2)} +
\psi_{0}^{(2)} )(a,t)\,\,\,.
\end{array}
\label{bc2d}
\EN
In these equations the intervals are labelled by the subscript of the fields
while their components by the upper indices. By using the notation $R_i$
and $T_i$ ($i=1,2$) for the reflection and the transmission amplitudes
relative to the defect line with strength $g_i$, it is easy to see that
eliminating the intermediate modes relative to the interval $I_0$, there is
a linear relationship between the modes of the fields in the intervals
$I_-$ and $I_+$ given by
\EQ
\left(\begin{array}{c}
A_-^{\dagger}(\beta) \\
A_+^{\dagger}(-\beta)
\end{array}
\right)
\,=\,
\left(\begin{array}{ll}
R(\beta,g_1,g_2,a) & T(\beta,g_1,g_2,a) \\
T(\beta,g_1,g_2,a) & R(\beta,g_1,g_2,a)
\end{array}
\right)\,
\left(\begin{array}{c}
A_-^{\dagger}(-\beta) \\
A_+^{\dagger}(\beta)
\end{array}
\right)
\,\, ,
\EN
where
\begin{eqnarray}
T(\beta,g_1,g_2,a) & = & \frac{T_1(\beta) T_2(\beta)}{1 - \eta(\beta,a)
R_1(\beta) R_2(\beta)}
\,\, ,
\nonumber\\
&&
\label{rt2}\\
R(\beta,g_1,g_2,a) & = & \, \frac{ R_1(\beta) + \eta(\beta,a) R_2(\beta)
[T_1^2(\beta) - R_1^2(\beta)]}{1 - \eta(\beta,a) R_1(\beta) R_2(\beta)}
\,\,\, .\nonumber
\end{eqnarray}
In the above expressions $\eta(\beta,a)$ is a pure phase given by
$\eta(\beta,a) = \exp[-2 i m a \sinh\beta]$.

The above amplitudes satisfy the unitarity and crossing equations
(\ref{unitarity}) and (\ref{crossing}). They describe the physical
situation of a particle coming from the interval $I_-$ with rapidity $\beta$
which hits the first defect line and, as result of this interaction,
it can be either reflected or transmitted. When it is reflected, it appears
as an asymptotic particle with rapidity $-\beta$ whereas when it is
transmitted it approaches the next defect and can again be reflected or
transmitted. As shown in Fig.\,6, these two types of process may be repeated
an arbitrary number of times at the two defect lines.

Due to the existence of the fixed points $g = \pm 2$ of a single defect line,
it is interesting to analyse some special limits of the expressions
(\ref{rt2}). To begin with, note that, at the values $g_1 = \pm 2$ where
$T_1 = 0$, the total transmission amplitude $T(\beta,g_1,g_2,a)$ vanishes as
well, whereas the reflection amplitude reduces to a pure phase given by
$R(\beta,\pm 2,g_2,a) = R_1(\beta,\pm 2)$. In this case, the first defect
acts as a pure reflecting surface which therefore completely screens
the presence of the second defect. The total transmission amplitude also
vanishes when $g_2=\pm 2$. Concerning the reflection amplitude, it becomes
a pure phase given by
\EQ
R(\beta,g_1,\pm 2,a)\,=\,\eta(\beta,a)\, R_2(\beta,\pm 2) \,
\frac{\sinh\beta (1+\eta^{-1}(\beta,a) \sin\chi_1) + i
\sin\chi (1-\eta^{-1}(\beta,a))}
{\sinh\beta (1+\eta(\beta,a) \sin\chi_1) - i
\sin\chi (1-\eta(\beta,a))} \,\,\,.
\EN
The total reflection process is now the result of an infinite sequence of
elementary transmission and reflection scatterings at the first defect line
combined with pure reflecting processes at the second defect line. Hence
it is not surprising that the final expression depends on both
$R_2(\beta,\pm 2)$ and the separation distance $a$.

Except for the values of $g$ when the defects behave as mirror surfaces, the
possibility for the fermion to go back and forth between the two defect lines
produces typical resonance phenomena which are illustrated for instance by
plotting the absolute value of $T(\beta,g_1,g_2,a)$. An example is shown in
Fig.\,7.

Finally, by taking the limit $a\rightarrow 0$, the two defect lines behave
as a single one but with an effective strength $g$ given by
\EQ
g\,=\, \frac{g_1 + g_2}{1 + g_1 g_2/4} \,\,\, .
\label{iter}
\EN
This composition law of the defect strengths is similar to the addition
of velocities in relativistic dynamics. The effective coupling constant $g$
has as critical values $g=\pm 2$ and reaches these limits when either $g_1$ or
$g_2$ are equal to $\pm 2$. This can be also seen by analysing the fixed
points of the composition law defined by the iterative map
\EQ
g_{n+1} \,=\, \frac{g_n + g}{1 + g_n g/4} \,\, ,
\EN
for some initial value $g$.

The natural generalization of the situation with two defect lines is to
consider a periodic array of defects all with equal strength $g$ and separated
by a distance $a$. The fermionic field satisfies in this case the equation
\EQ
\left[i\gamma^{\mu}\partial_{\mu} - m - g\,\sum_{n=-\infty}^{\infty}
\delta(x + n a)\right] \Psi(x,t)\,=\, 0 \,\, ,
\EN
and admits the decomposition
\EQ
\Psi(x,t) \,=\,\sum_{n=-\infty}^{\infty}
\theta(x - n a) \theta(-x + (n+1) a) \Psi_n(x,t) \,\, ,
\EN
with $\Psi_n(x,t)$ solutions of the free Dirac equation. The dynamics of the
model is entirely encoded into an infinite set of linear equations relative
to the boundary conditions between the interval $n a$ and $(n-1) a$, i.e.
\EQ
\begin{array}{lll}
(\psi_{n-1}^{(2)} - \psi_{n}^{(2)})(na,t) & = & \frac{g}{2} (\psi_{n-1}^{(1)} +
\psi_{n}^{(1)})(na,t) \,\,\, ;\\
(\psi_{n-1}^{(1)} - \psi_{n}^{(1)})(na,t) & = & \frac{g}{2} (\psi_{n-1}^{(2)} +
\psi_{n}^{(2)} )(na,t)\,\,\, .
\end{array}
\label{bcn}
\EN
The simplest way to solve these equations is to employ a relativistic
generalization of the Bloch theorem \cite{CMB}, i.e to associate a wave vector
$k$ to the spinor field $\Psi$ such that
\EQ
\Psi(x+a,t)\,=\,e^{i k a} \,\Psi(x,t) \,\,\, .
\label{bloch}
\EN
Equivalently,
\EQ
\Psi_n(na,t) \,=\,e^{i k a} \,\Psi_{n-1}((n-1) a, t) \,\,\,.
\label{bloch2}
\EN
The wave vector $k$ can always be confined to the first Brillouin zone
$-\pi/a \leq k \leq \pi/a$. Plugging (\ref{bloch2}) into eqs. (\ref{bcn}),
the resulting system is compatible provided that the equation
\EQ
\cos k a \,=\, \frac{1}{\cos\chi} \left[
\cos(m a \sinh\beta) - \sin\chi \frac{\sin(m a \sinh\beta)}{\sinh\beta}\right]
\label{bande}
\EN
is valid. This equation gives rise to a band structure in the energy levels
of the Majorana fermion of the Ising model, completely analogous to the
periodic potentials considered in condensed matter physics. In fact, eq.
(\ref{bande}) can be satisfied for real $k$ if and only if the right hand
side of the equation is less than unity. Consequently, there will be
allowed and forbidden regions of $\beta$ and the corresponding spectrum
of the energy, given by $E = m \cosh\beta$, consists of a family of
energy bands. A characteristic form of the spectrum is plotted in Fig.\,8.
For the pure reflecting values $g=\pm 2$, the above equation
reduces to the quantization condition of the rapidity variable $\beta$
\EQ
\sinh\beta \,=\, \pm \tan(m a \sinh\beta) \,\, ,
\EN
which arises by considering the fermionic field defined in a strip of width
$a$ with fixed ($+$) or free ($-$) boundary conditions at the edge of the
interval.

\subsection{Bosonic Theory}

The discussion of the bosonic theory largely follows the previous one and
eqs.\,(\ref{rt2}) is valid as it stands on the condition that we insert the
bosonic amplitudes instead. Also in this case there are typical resonance
phenomena produced by the trapping of the bosonic particle between the two
defect lines. There is however a significant difference with
respect to the fermionic case and this concerns the composition law
relative to two defect lines with a separation $a\rightarrow 0$. In this limit,
the two defect lines behave as a single one with an effective strength
$g$ given by
\EQ
g\,=\, g_1 + g_2 \,\,\, .
\EN
Due to the peculiar properties of the bosonic system discussed in Section 4,
this composition law implies that a system with two defect lines in the
limit $a\rightarrow 0$ may become unstable although each of
the defect lines taken individually does not present any instability property.
Viceversa, one can obtain a well-defined bosonic system as a result of the
limit $a\rightarrow 0$ of a system which presents instability properties
at one defect line and resonance states at the other.

Taking the limit $g_1\rightarrow +\infty$, the first defect line becomes
a pure reflecting surface and the total transmission amplitude vanishes.
In this case the reflection amplitude reduces to $R(\beta,+\infty,g_2,a)=-1$.
The total transmission amplitude also vanishes when the second defect line
acts as a pure reflecting surface. The corresponding reflection amplitude
is a pure phase given by
\EQ
R(\beta,g_1,+\infty,a)\,=\,-\eta(\beta,a)\,
\frac{\sinh\beta - i \frac{g}{4m}(1-\eta^{-1}(\beta,a))}
{\sinh\beta + i \frac{g}{4m}(1-\eta(\beta,a))}\,\,\,.
\EN

As in the fermionic case, the presence of an infinite periodic array of defect
lines of strength $g$ and separation $a$ gives rise to a band structure
described by a Kronig-Penney type equation
\EQ
\cos k a \,=\, \cos(m a \sinh\beta) + \frac{g}{m}\,
 \frac{\sin(m a \sinh\beta)}{\sinh\beta} \,\,\, .
\EN
The pure reflective case $g\rightarrow +\infty$ gives rise to the
quantization condition
\EQ
m a \sinh\beta \,=\, \pi n \,\,\, , \hspace{3mm} (n=0,\pm 1,\ldots)
\EN
relative to the bosonic field in a strip of width $a$ with fixed boundary
conditions $\varphi(0,t) = \varphi(a,t) = 0$ at the end points of the interval.

\resection{Correlation Functions}

In the bulk, the scattering theory and the bootstrap approach -in addition
to yield a clear understanding on the physical content of the continuum
limit of the integrable statistical models- have the added advantage of
providing a powerful method for the computation of the correlation functions of
the order parameters. Once the bulk $S$-matrix is known, there are
well-defined techniques for computing matrix elements of the local fields
$\phi_i(x,t)$ on the set of asymptotic states
$<\beta_1,\ldots \beta_n \mid \phi_i(x,t)\mid \beta_{n+1},\ldots \beta_m>$
and for reconstructing their correlation functions through spectral
representation method based on the completeness relation of the asymptotic
states. This program, known as Form Factor Approach, was originally proposed
in \cite{KW,Smirnov} and through this method many QFT have been recently
solved [9-15]. In this section we will not go into the details of the Form
Factor Approach in the bulk, which can be found in the aforementioned
literature. We do however intend to prove that the spectral methods are
also suitable for computing the correlation functions of the Ising model and
the bosonic theory in the presence of a defect line.

The easiest way to approach the problem is to use a formalism which takes
full advantage of the solution of the theory in the bulk. To this aim, it is
convenient to interchange the original role of the $x$ and the $t$ axes
by the transformation $x\rightarrow -i t$, $t\rightarrow i x$. The new space
has a Minkowski structure with the defect line placed now at $t=0$.
In this new geometry, the space of the states is the same as in the bulk, and
therefore, even in the presence of the defect line, the local operators
$\phi_i$ can be completely characterized by their known form factors. The
presence of the defect line can be taken into account by defining an operator
${\cal D}$ placed at $t=0$, acting on the bulk states. This operator plays
the role of the $S$-matrix of the problem, and therefore, standard formulas
of QFT allow the correlation functions to be expressed as \cite{Landau}
\EQ
<\Phi_1(x_1,t_1) \ldots \Phi_n(x_n,t_n)> \,=\, \frac{<0\mid T
[\phi_1(x_1,t_1) \ldots {\cal D} \ldots \phi_n(x_n,t_n) ]\mid 0>}
{<0\mid {\cal D}\mid 0>}
\,\,\,.
\label{correlation}
\EN
In the above formula, $\Phi_i(x_i,t_i)$ are the fields in the Heisenberg
representation, i.e. the representation where the time evolution is ruled by
the exact Hamiltonian of the problem, including the defect interaction. On the
other hand, $\phi_i(x_i,t_i)$ are the field operators of the bulk theory and,
as such, their time evolution operator is the bulk Hamiltonian\footnote{An
equivalent way to look at eq.\,(\ref{correlation}) is to consider a transfer
matrix approach in the euclidean space. The transfer matrix may be written as
${\cal T} = \exp [-H_B t]$ for all $t$ but $t=0$, where it is placed
the defect line. Hence ${\cal D}$ in (\ref{correlation}) can be interpreted
as the continuum limit of the transfer matrix operator which connects the
states below and above the defect line.}.
The main advantage of eq.\,(\ref{correlation}) is that, using the completeness
relation of the bulk states, its right hand side can be entirely
expressed in terms of the Form Factors of the bulk fields and the matrix
elements of the operator ${\cal D}$ which are determined as follows.

The defect operator ${\cal D}$ encodes all information relative to the
physical processes which take place at the defect line. To examine them,
we have to initially realize that the first effect of the interchange of
the $x$ and the $t$ axes consists in an analytic continuation of the original
rapidity $\beta \rightarrow \left(i\frac{\pi}{2} - \beta\right)$, the reason
being that, to preserve the Minkowski structure in the new set of axes, we
have to interchange correspondingly the momentum and the energy role. The
rapidities are now measured as in Fig.\,9. For convenience, it is
useful to introduce the new transmission and reflection amplitudes, given
by
\EQ
\hat T(\beta)\,=\,T\left(i\frac{\pi}{2} - \beta\right)
\,\,\, ,
\hspace{5mm}
\hat R(\beta)\,=\,R\left(i \frac{\pi}{2} - \beta\right)
\,\,\,.
\EN
They enter the expression of the simplest matrix elements of the operator
${\cal D}$, given by ${\cal D}_{1,1}=<\beta\mid{\cal D}\mid\theta>$,
${\cal D}_{2,0} = <\beta_1,\beta_2\mid {\cal D}\mid 0>$ and
${\cal D}_{0,2} = <0\mid {\cal D}\mid \beta_1,\beta_2>$. For the fermionic
and the bosonic theory analysed in the previous sections, the first matrix
element is easily computed by resumming the perturbative series with the defect
interaction now localized at $t=0$ and the result is
\EQ
<\beta\mid{\cal D}\mid\theta>\,=\, 2\pi\,\hat T(\beta)\,\delta(\beta - \theta)
\,\,\,.
\EN
By the same means, for the other two matrix elements, we have respectively
\EQ
<\beta_1,\beta_2\mid {\cal D}\mid 0>\,=\,2\pi\,\hat R(\beta_1) \,
\delta(\beta_1 + \beta_2)
\,\, ,
\label{creation}
\EN
and
\EQ
<0\mid {\cal D}\mid \theta_1,\theta_2>\,=\,2\pi\,
\hat R(\theta_1) \,
\delta(\theta_1 + \theta_2)
\,\,\,.
\label{annihilation}
\EN
Hence, $\hat T(\beta)$ describes the process where a particle with rapidity
$\beta$ hits the defect line and is transmitted through it, keeping the same
value of the rapidity (Fig.\,9.a). On the contrary, $\hat R(\beta)$ may be
interpreted as the amplitude for the creation or the annihilation of a pair
of particles with equal and opposite rapidity $\beta$ (Fig.\,9.b). These three
processes are compatible with the dynamics of the model because in a situation
where the defect line is placed at $t=0$, the processes are constrained by the
conservation of the momentum but not of the energy.

For the general matrix elements of the operator ${\cal D}$, we can exploit
the factorization property of the scattering theory and write down a
set of recursive equations which involve the elementary two-body
interactions considered above. For the bosonic case, the recursive equations
are expressed by
\begin{eqnarray}
&& <\beta_1,\ldots,\beta_i,\ldots,\beta_m,\beta\mid{\cal D}\mid
\theta_1,\ldots \theta_n>
\,=\, \nonumber\\
&& \hspace{1mm}
= 2\pi\,\sum_{i=1}^m \hat R(\beta) \delta(\beta+\beta_i)
\,<\beta_1,\ldots,\beta_{i-1},\beta_{i+1},\ldots \beta_m\mid{\cal D}\mid
\theta_1,\ldots \theta_n> + \\
&& \hspace{1mm} +\,2\pi\,\sum_{j=1}^n \hat T(\beta) \,\delta(\beta-\theta_j)\,
<\beta_1,\ldots \beta_m\mid
{\cal D}\mid \theta_1,\ldots \theta_{j-1},\theta_{j+1},\ldots,\theta_n> \,\,\,;
\nonumber
\end{eqnarray}

\vspace{1cm}

\begin{eqnarray}
&& <\beta_1,\ldots,\beta_m,\mid{\cal D}\mid
\theta_1,\ldots \theta_n,\theta>
\,=\, \nonumber\\
&& \hspace{1mm} = 2\pi\,\sum_{i=1}^n \hat R(\theta) \delta(\theta+\theta_i)
\,<\beta_1,\ldots, \beta_m\mid{\cal D}\mid
\theta_1,\ldots,\theta_{i-1},\theta_{i+1},\ldots, \theta_n> + \\
&& \hspace{1mm} + 2\pi\,\sum_{j=1}^m \hat T(\theta) \delta(\theta-\beta_j)\,
<\beta_1,\ldots,\beta_{j-1},\beta_{j+1},\ldots, \beta_m\mid
{\cal D}\mid \theta_1,\ldots,\theta_n>\,\,\,.
\nonumber
\end{eqnarray}
For the fermionic case, taking into account the anti-commutation relations
of the fields, they can be written as
\begin{eqnarray}
&& <\beta_1,\ldots,\beta_i,\ldots,\beta_m,\beta\mid{\cal D}\mid
\theta_1,\ldots \theta_n>
\,=\, \\
&& \hspace{1mm} = 2\pi\,\sum_{i=1}^m
(-1)^{m+1-i}\,\hat R(\beta) \delta(\beta+\beta_i)
\,<\beta_1,\ldots,\beta_{i-1},\beta_{i+1},\ldots \beta_m\mid{\cal D}\mid
\theta_1,\ldots \theta_n> + \nonumber \\
&& \hspace{1mm} + 2\pi\,
\sum_{j=1}^n (-1)^j\,\hat T(\beta) \,\delta(\beta-\theta_j)\,
<\beta_1,\ldots \beta_m\mid
{\cal D}\mid \theta_1,\ldots \theta_{j-1},\theta_{j+1},\ldots,\theta_n>\,\,\,;
\nonumber
\end{eqnarray}
\begin{eqnarray}
&& <\beta_1,\ldots,\beta_m,\mid{\cal D}\mid
\theta_1,\ldots \theta_n,\theta>
\,=\, \nonumber \\
&& \hspace{1mm} = 2\pi\,\sum_{i=1}^n
(-1)^{i}\hat R(\theta) \delta(\theta+\theta_i)
\,<\beta_1,\ldots, \beta_m\mid{\cal D}\mid
\theta_1,\ldots,\theta_{i-1},\theta_{i+1},\ldots, \theta_n> + \\
&& \hspace{1mm} + 2\pi\,\sum_{j=1}^m (-1)^{m-j}
\hat T(\theta) \delta(\theta-\beta_j)\,
<\beta_1,\ldots,\beta_{j-1},\beta_{j+1},\ldots, \beta_m\mid
{\cal D}\mid \theta_1,\ldots,\theta_n>\,\, ,
\nonumber
\end{eqnarray}
These recursive equations can be graphically represented as in Fig.\,10 and
express the exact resummation of the perturbative series associated to
the scattering matrix elements $<m\mid{\cal D}\mid n>$. Since the particles
are created or destroyed in couples, the non-vanishing matrix element
$<m\mid{\cal D}\mid n>$ are only those with $m-n = 0$ (mod $2$). They are
proportional to $<0\mid {\cal D}\mid 0>$ (which, for convenience, is set
equal to $1$) and the recursive equations permit to express all of them
in terms of the elementary matrix elements ${\cal D}_{1,1}, {\cal D}_{2,0}$ and
${\cal D}_{0,2}$ as previously determined.

A useful method for solving the recursive equations is to introduce a
generating functional of the matrix elements of ${\cal D}$ by the formula
\EQ
{\cal G}(\eta,\gamma)\,=\,
\exp\left\{
\int d\beta \left(
\frac{\hat R(\beta)}{2} \left[\eta(-\beta) \eta(\beta) +
\gamma(\beta) \gamma(-\beta) \right] +
\hat T(\beta) \eta(\beta) \gamma(\beta)\right)\right\}
\,\,\,.
\label{defoper}
\EN
${\cal G}$ depends on the two currents $\eta(\beta)$ and $\gamma(\beta)$,
which commute or anti-commute, depending on whether we are considering the
bosonic theory or the fermionic one. The matrix elements of ${\cal D}$ are
then given by
\EQ
<\beta_1,\ldots,\beta_m,\mid{\cal D}\mid \theta_1,\ldots \theta_n> \,=\,
(2\pi)^{\frac{m+n}{2}}\, \left.
\frac{\partial}{\partial \gamma(\theta_n)}\ldots
\frac{\partial}{\partial \gamma(\theta_1)}
\frac{\partial}{\partial \eta(\beta_1)}\ldots
\frac{\partial}{\partial \eta(\beta_m)}\,{\cal G}
\right|_{\eta=\gamma=0}
\,\,\,.
\label{generating}
\EN.

We are now in the position to compute correlation functions of local operators
of the Ising model and the bosonic theory with a line of defect.
Note that in computing the left hand side of eq. (\ref{correlation}) we should
consider two different cases, namely: ({\em a}) the case where some of the
operators $\Phi_i$ are in the upper half-plane and the others are in the lower
one, or ({\em b}) the case where the operators $\Phi_i$ are all in one
semi-plane, for example the upper one. In the former case, one has to use the
general matrix elements $<i\mid{\cal D}\mid j>$, and consequently both
transmission and reflection amplitudes will enter the final expression of the
correlation functions. In the latter case, on the contrary, the
correlation functions will depend only on the reflection amplitudes
$\hat R(\beta)$ because, in virtue of the time ordering in eq.
(\ref{correlation}), the defect operator ${\cal D}$ is in this case the last
in the row and so, it acts directly on the vacuum state $\mid 0>$. Hence, the
only matrix elements which enter the computation are
${\cal D}_{i,0}=<i\mid{\cal D}\mid 0>$. Those describe the creation of the
particle pairs and therefore only depend on $\hat R(\beta)$.

In the remaining part of this section, using the form factors of the Ising
model determined in \cite{KW,ZCM}, and the matrix elements of
the defect operator we compute some correlation functions of this model in
the presence of the defect line\footnote{All correlation functions will be
computed in the euclidean space obtained by the analytic continuation
$t\rightarrow i t$.}. The simplest one is the one-point function of the
energy operator $\epsilon(x,t)$ which can be computed through the formula
\EQ
\epsilon_0(t)\,=\,\sum_{n=0}^{\infty} <0\mid\epsilon(x,t)\mid n>
<n\mid{\cal D}\mid 0>\,\,\,.
\label{e01}
\EN
The energy operator couples the vacuum only to the two particle state,
as can be easily checked by the fermionic representation of this operator,
and for its matrix element we have
\begin{eqnarray}
&& <0\mid \epsilon(x,t)\mid \beta_1,\beta_2>\,=\,2\pi i \,m \,
\sinh\frac{\beta_1-\beta_2}{2} \times
\label{ffe} \\
&& \hspace{3mm}\times \,\exp\left[-m t\, (\cosh\beta_1 + \cosh\beta_2) +
i m x \,(\sinh\beta_1 + \sinh\beta_2)\right]\,\, ,\nonumber
\end{eqnarray}
Hence the above sum (\ref{e01}) consists of only one term (Fig.\,11) and using
eq.\,(\ref{creation}), it can be expressed as
\EQ
\epsilon_0(t)\,=\,m\,\sin\chi\,\int_0^{\infty}d\beta
\frac{\sinh^2\beta}{\cosh\beta - \sin\chi}\,e^{-2mt\cosh\beta} \,\,\, .
\label{energy1}
\EN
The one-point function does not depend on $x$, as it can be equivalently
argued by translation invariance along this axis. The above integral reduces
to closed expressions in terms of Bessel functions when the defect line acts
as pure reflecting surface. In the case of fixed boundary conditions, we have
\EQ
\epsilon(t)\,=\,-m\,\left[K_1(2mt) - K_0(2mt)\right] \,\, ,
\label{up}
\EN
whereas for free boundary conditions
\EQ
\epsilon(t)\,=\,m\,\left[K_1(2mt) + K_0(2mt)\right] \,\, ,
\label{down}
\EN
In the general case, the one-point function interpolates between the
two curves. The critical exponent of the energy operator in the presence of the
defect line can be extracted by looking at the ultraviolet limit
$t\rightarrow 0$ of its one-point function. For this limit we have
\EQ
\epsilon_0(t) \sim\frac{\sin\chi}{2t} \,\,\,.
\label{ultrav}
\EN
{}From this expression, we see that the defect line does not influence the
critical exponent of the energy operator, which is the same as in the bulk,
but rather enters the universal amplitude of the one-point function.
For the pure reflecting case, the universal amplitudes coincide with those
calculated in \cite{CL}.

The relationship between the coupling constant in the continuum theory
and in the discrete formulation can be extracted by comparing eq.
(\ref{ultrav}) with the analogous lattice computation, which reads \cite{Ko}
\EQ
\epsilon_0(t) \sim \frac{\tanh2(J -\tilde J)}{2t}
\,\,\,.
\EN
Hence, we have the following identification
\EQ
\sin\chi \,=\, \tanh2(J -\tilde J) \,\,\, .
\label{lat-cont}
\EN
In addition to the one-point function of the energy operator, it is also
interesting to compute its two-point function. To simplify calculations,
it is convenient to define the function
\EQ
F(x,t)\,=\,\frac{1}{2} \,\int_{-\infty}^{\infty} d\beta \,
\frac{\exp[-t \cosh\beta + i x \sinh\beta]}{\cosh\beta - \sin\chi}\,\,\,.
\label{auxiliar}
\EN
Let us initially consider the situation where the energy density operators are
on opposite sides of the defect line, i.e. $t_2 >0$ and $t_1 <0$.
The relevant expression in this case is given by\footnote{To simplify the
notation, in the sequel we denote the couple of coordinate $(x_i,t_i)$ simply
by $\rho_i$.}
\EQ
G_1(\rho_1,\rho_2)\,=\,
\sum_{i,j} <0\mid \epsilon(\rho_2)\mid i><i\mid{\cal D}\mid j><j\mid
\epsilon(\rho_1)\mid 0> \,\,\,.
\label{twoacross}
\EN
As before, the above series terminate. To explicitly evaluate it, in addition
to the matrix elements ${\cal D}_{2,0}$ and ${\cal D}_{0,2}$, we also need the
matrix element ${\cal D}_{2,2}$ given by
\begin{eqnarray}
<\beta_1,\beta_2\mid {\cal D}\mid \theta_1,\theta_2> & = &\,
(2\pi)^2 \left[ \hat R(\beta_1)\, \hat R(\theta_1) \delta(\beta_1+\beta_2)
\delta(\theta_1+\theta_2) \, + \right. \nonumber  \\
& & \hspace{10mm} +\, \hat T(\beta_1)\, \hat T(\beta_2)
\delta(\beta_1 - \theta_1) \,
\delta(\beta_2 - \theta_2) \, +
\label{d22} \\
& & \hspace{10mm}  -\, \left. \hat T(\beta_1)\,
\hat T(\beta_2) \,\delta(\beta_1 - \theta_2) \,
\delta(\beta_2 - \theta_1) \right] \,\,\,.
\nonumber
\end{eqnarray}
With the notation $t\equiv t_2 - t_1$ and $x \equiv x_2 - x_1$, eq.
(\ref{twoacross}) can be expressed as
\begin{eqnarray}
G_1(\rho_1,\rho_2) & = & \cos^2\chi
\,\left[\left(\frac{\partial^2}{\partial x \partial t} F(x,t)\right)^2 +
\left(\frac{\partial^2}{\partial t^2} F(x,t)\right)^2 -
\left(\frac{\partial}{\partial t} F(x,t)\right)^2\right] +
\nonumber \\
&& \hspace{1mm} + \,\epsilon_0(t_1) \,\epsilon_0(t_2)
\,\,\,.
\label{facross}
\end{eqnarray}
When the defect line acts as a pure reflecting surface, all fluctuations
across it are suppressed and this formula correctly reduces to the vacuum
expectation values of the energy densities.

Let us consider now the situation where the two energy operators are on the
same side of the defect line, with $t_2 \geq t_1 >0$. For convenience,
let us introduce the notation $t\equiv t_2 - t_1$,
$\overline t\equiv t_2 + t_1$, $x \equiv x_2 - x_1$ and
$r\equiv \sqrt{x^2+t^2}$. The two point function can be written in this case
as
\EQ
G_2(\rho_1,\rho_2)\,=\,\sum_{i,j} <0\mid\epsilon(\rho_2)\mid i><i\mid
\epsilon(\rho_1)\mid j><j\mid {\cal D}\mid 0> \,\,\,.
\label{twoside}
\EN
There are only a finite number of non-vanishing matrix elements of the
energy density and therefore the series truncates. It can be written as
\EQ
G_2(\rho_1,\rho_2)\,=\,I_{1} + I_2 + I_3 \,\, ,
\EN
where
\begin{eqnarray*}
I_1 &=& \frac{1}{2!} \int_{-\infty}^{+\infty}
\frac{d\beta_1}{2\pi} \frac{d\beta_2}{2\pi}
<0\mid\epsilon(\rho_2)\mid\beta_1,\beta_2><\beta_1,\beta_2\mid
\epsilon(\rho_1)\mid 0><0\mid {\cal D}\mid 0> \\
I_2 &=& \frac{1}{2!\,2!} \int_{-\infty}^{+\infty}
\frac{d\beta_1}{2\pi}\ldots \frac{d\beta_4}{2\pi}
<0\mid\epsilon(\rho_2)\mid \beta_1,\beta_2><\beta_1,\beta_2\mid
\epsilon(\rho_1)\mid \beta_3,\beta_4><\beta_3,\beta_4
\mid {\cal D}\mid 0> \\
I_3 &=& \frac{1}{2!\,4!} \int_{-\infty}^{+\infty}
\frac{d\beta_1}{2\pi}\ldots \frac{d\beta_6}{2\pi}
<0\mid\epsilon(\rho_2)\mid \beta_1,\beta_2><\beta_1,\beta_2\mid
\epsilon(\rho_1)\mid \beta_3,..,\beta_6><\beta_3,..,\beta_6\mid {\cal D}\mid 0>
\end{eqnarray*}
$I_1$ coincides with the two-point function of the energy operator in the bulk,
\[
I_1\,=\, m^2\left[\left(\frac{\partial }{\partial x}
K_0(mr)\right)^2 + \left(\frac{\partial}{\partial t} K_0(mr)\right)^2
- (K_0(mr))^2 \right]\,\,\,.
\]
The quantities which appear in $I_2$ and $I_3$ are the higher matrix elements
of the energy density (which may be directly computed by the fermionic
representation of this operator, $\epsilon = i\overline\Psi \Psi$) and the
matrix elements of the defect operator ${\cal D}$, given by
(\ref{generating}). Considering that the computation of these quantities is
lengthy but straightforward, we shall only present the final result
\begin{eqnarray*}
I_2 &=& 2\,m^2\,\sin\chi \left[
\left(\frac{\partial}{\partial x} K_0(r) \right)
\left(\frac{\partial}{\partial x} F(x,\overline t)\right) -
K_0(r) \left(\frac{\partial^2}{\partial x^2} F(x,\overline t)\right)
\right]\,\, ,
\\
I_3 &=& m^2\,\sin^2\chi
\left[\left(\frac{\partial}{\partial x} F(x,\overline t)\right)^2 -
\left(\frac{\partial^2}{\partial x^2} F(x,\overline t)\right)^2 - \,
\left(\frac{\partial^2}{\partial x \partial \overline t} F(x,\overline t)
\right)^2 \right] +
\\
& & \hspace{3mm} +\, \epsilon_0(t_1) \epsilon_0(t_2) \,\,\,. \nonumber
\end{eqnarray*}
Returning to eq.\,(\ref{twoside}), the two-point function can be cast in the
form
\begin{eqnarray}
G_2(\rho_1,\rho_2) & & = \epsilon_0(t_1) \epsilon_0(t_2) +
\left[\frac{\partial}{\partial x} K_0(r) + \sin\chi\,
\frac{\partial}{\partial x}
F(x,\overline t)\right]^2 + \left[\frac{\partial}{\partial t} K_0(r)
\right]^2 \nonumber \\
& & - \left[\sin\chi \,\frac{\partial}{\partial x \partial\, \overline t}
F(x,\overline t)\right]^2 - \left[K_0(r) + \sin\chi\, \frac{\partial^2}
{\partial x^2} F(x,\overline t)\right]^2
\,\,\,. \label{fside}
\end{eqnarray}
It is now easy to verify that the expressions (\ref{facross}) and
(\ref{fside}) coincide with those obtained in the lattice calculation
\cite{Ko}.

As our last example, we discuss the one-point function of the magnetization
operator $\sigma(\rho)$ in the low temperature phase in the presence of
the defect line. It can be calculated through the formula
\EQ
\sigma_0(t)\,=\,\sum_{n=0}^{\infty} <0\mid \sigma(\rho)\mid n>
<n\mid {\cal D}\mid 0> \,\,\,.
\label{magvac}
\EN
The magnetization operator couples the vacuum to all states with an even
number of particles and its form factors are given by \cite{KW,ZCM}
\EQ
<0\mid \sigma(0,0)\mid \beta_1,\ldots \beta_{2n} > \,=
\, (-i)^n \prod_{i<j} \tanh\frac{\beta_i-\beta_j}{2}
\,\,\,.
\EN
Since the matrix elements of ${\cal D}$ in (\ref{magvac}) are different from
zero only for pairs of particles of opposite momentum, we are lead to consider
the matrix elements of the magnetization operator given by
$<0\mid\sigma(0)\mid -\beta_1,\beta_1,\ldots,-\beta_n,\beta_n>$. They can be
conveniently written as
\EQ
<0\mid \sigma(0,0)\mid -\beta_1,\beta_1,\ldots -\beta_n,\beta_n>\,=\,
i^n \left(\prod_{i=1}^n \tanh\beta_i\right)
\times {\makebox det}\, W(\beta_i,\beta_j)
\,\, ,
\EN
where $W(\beta_i,\beta_j)$ is the $n\times n$ matrix given by
\EQ
W(\beta_i,\beta_j)\,=\, \left(\frac{2\,\sqrt{\cosh\beta_i\cosh\beta_j}}
{\cosh\beta_i+\cosh\beta_j}
\right)\,\,\,.
\EN
Hence, the one-point function is the sum of an infinite number of terms shown
in Fig.\,12 and it can be expressed as a Fredholm determinant
\begin{eqnarray}
\sigma_0(t) & = & \sum_{n=0}^{\infty} \frac{1}{n!} \int_{-\infty}^{\infty}
d\beta_1\ldots d\beta_n \,\left(\prod_{k=0}^n i\,\tanh\beta_k \,
\hat R(\beta_k)\,e^{-2mt \cosh\beta_k} \right)
 \,{\makebox det} W(\beta_i,\beta_j) \,= \nonumber \\
& & =\,{\makebox Det}\,(1 + z\,{\cal W}) \,\,\,.
\label{mu}
\end{eqnarray}
The explicit form of the kernel is given by
\EQ
{\cal W}(\beta_i,\beta_j,\chi)\,=\,\frac{E(\beta_i, mt,\chi)
E(\beta_j,mt,\chi)}{\cosh\beta_i + \cosh\beta_j}\,\, ,
\label{kernel}
\EN
where
\EQ
E(\beta,mt,\chi) \,= \,\sinh\beta\,e^{-mt \cosh\beta}\,
(\cosh\beta - \sin\chi)^{-1/2}
\hspace{3mm},\hspace{5mm}
z \,=\,\frac{\sin\chi}{2\pi} \,\,\, .
\EN
In terms of the eigenvalues of the integral operator and their multiplicity,
$\sigma_0(t)$ can be also expressed as
\EQ
\sigma_0(t)\,=\,\prod_{i=1}^{\infty} \left(1 + z\,\lambda_i\right)^{a_i}
\label{eigenvalues}
\EN
As far as $mt$ is finite, the kernel is square integrable and therefore all
results valid for bounded symmetric integral operators apply (see, for instance
\cite{Integral}). In particular, for $mt\rightarrow \infty$, $\sigma_0(t)$
falls off exponentially to the bulk vacuum expectation value. However,
when $mt \rightarrow 0$, the integral operator becomes unbounded.
The multiplicity of the eigenvalues grows logarithmically as
$a \sim \frac{1}{\pi} \ln\frac{1}{mt}$ whereas the eigenvalues become
dense in the interval $(0,\infty)$ according
to the distribution
\EQ
\lambda(p)\,=\,\frac{2\pi}{\cosh\pi p} \,\,\, .
\EN
Hence, for the critical exponent of the magnetization operator, defined by
\EQ
\sigma_0(t)\, \sim \, \frac{C}{\,\,\,(2t)^{x_{\sigma}}} \,\,\, ,
\,\,\, t\rightarrow 0
\,\, ,
\EN
we have
\EQ
x_{\sigma}(\chi)\,= -\,\frac{1}{\pi}\,
\int_0^{\infty} dp \,\ln\left(1 + \frac{2\pi z}{\cosh p}\right)
\,=\,- \frac{1}{8} +\frac{1}{2\pi^2} \arccos^2(-\sin\chi)\,\,\, .
\EN
This expression agrees with the lattice calculations \cite{Bariev,Mc} and
since it depends on the coupling constant, it explicitly shows the
non-universality of the model.

\resection{Conclusion}

The main purpose of this paper was to prove that the bootstrap approach
can be successfully extended to integrable models with linear inhomogeneities
and that the computation of the correlation functions for those systems can
be achieved by means of a suitable generalization of the Form Factor
techniques. We have analysed the general situation in which translation
invariance is broken by the presence of defect lines allowing reflection and
transmission processes. While at the moment it is still an open problem to
see whether there are other solutions of the Transmission-Reflection equations
in addition to the fermionic and the bosonic theories analysed in the text, it
is worth to stress that the method for the computation of the correlation
functions exposed in Sec.\,6 is expected to work without limitation in the
pure reflecting case. This corresponds to the boundary field theories which
have recently received a lot of attention in view of their potential
application to a wide class of physical situations.

\vspace{5mm}
\noindent
{\em Acknowledgements}. We are grateful to R. Iengo and M. Nolasco
for useful discussions.

\pagestyle{empty}

\newpage

\hs

\vspace{25mm}

{\bf Figure Captions}

\vspace{1cm}

\begin{description}
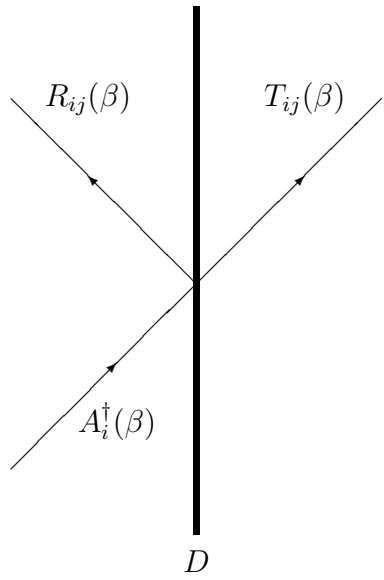
\item [Figure 1]. Reflection and Transmission Amplitudes.
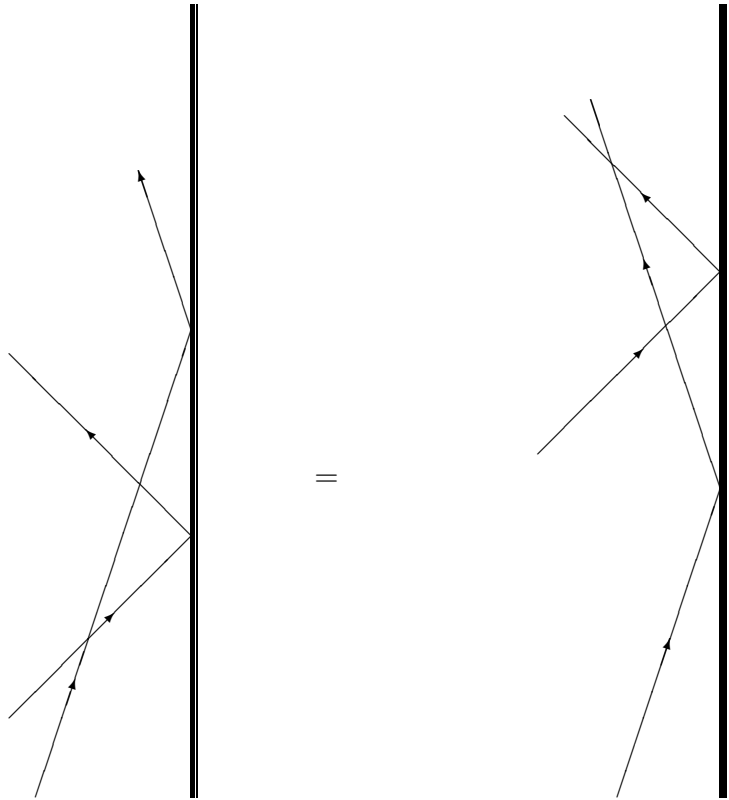
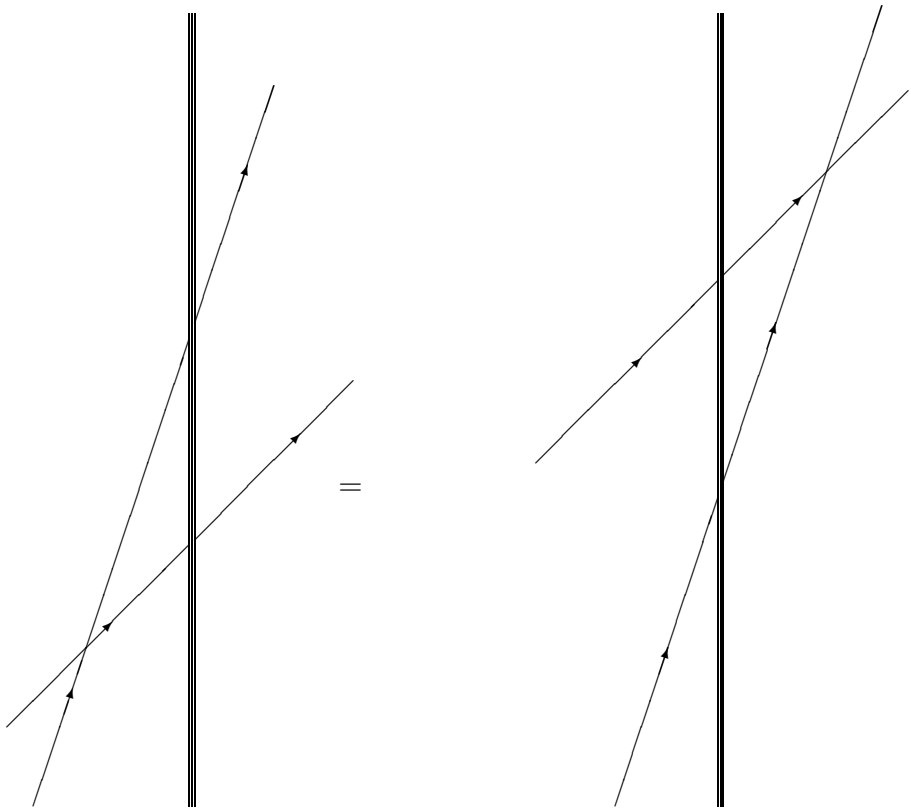
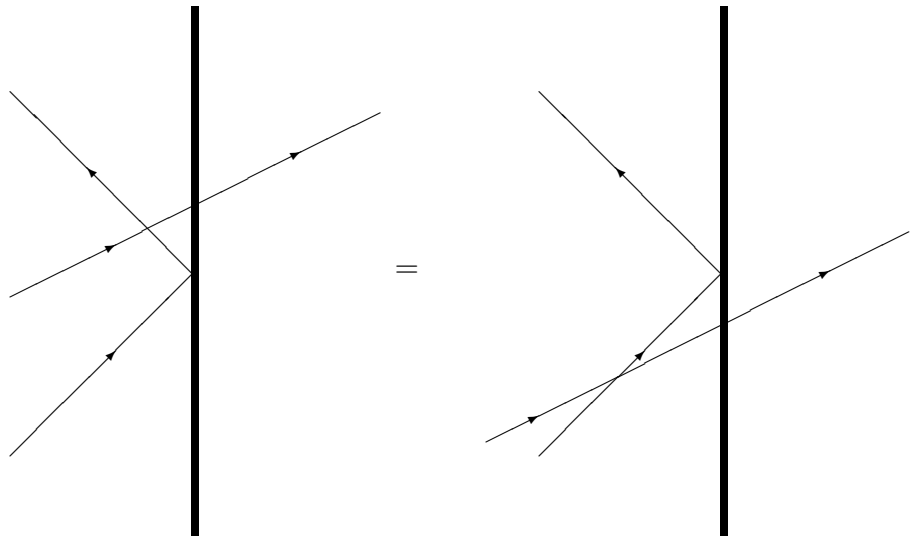
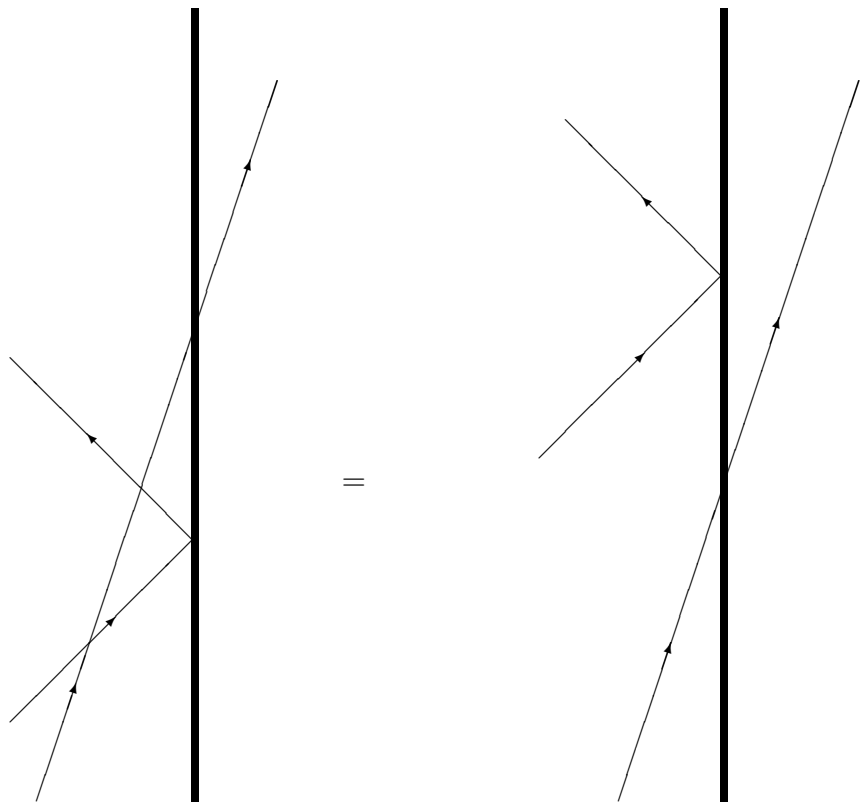
\item [Figure 2]. Reflection-Transmission Equations.
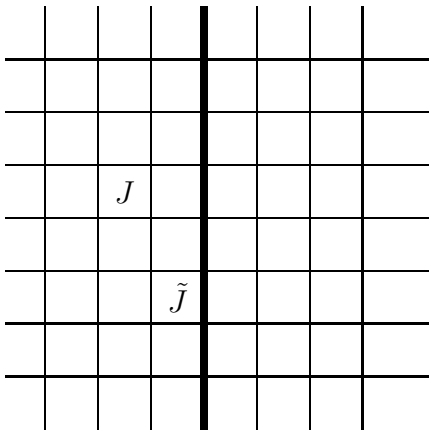
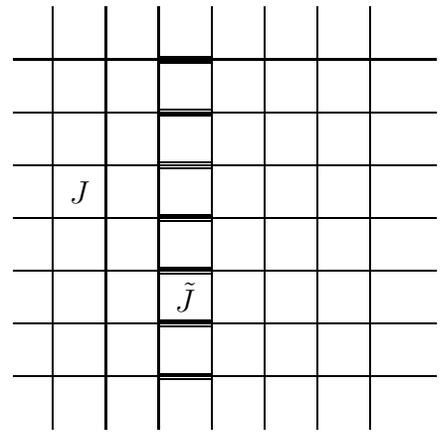
\item [Figure 3]. Chain and ladder geometry of a defect line lattice.
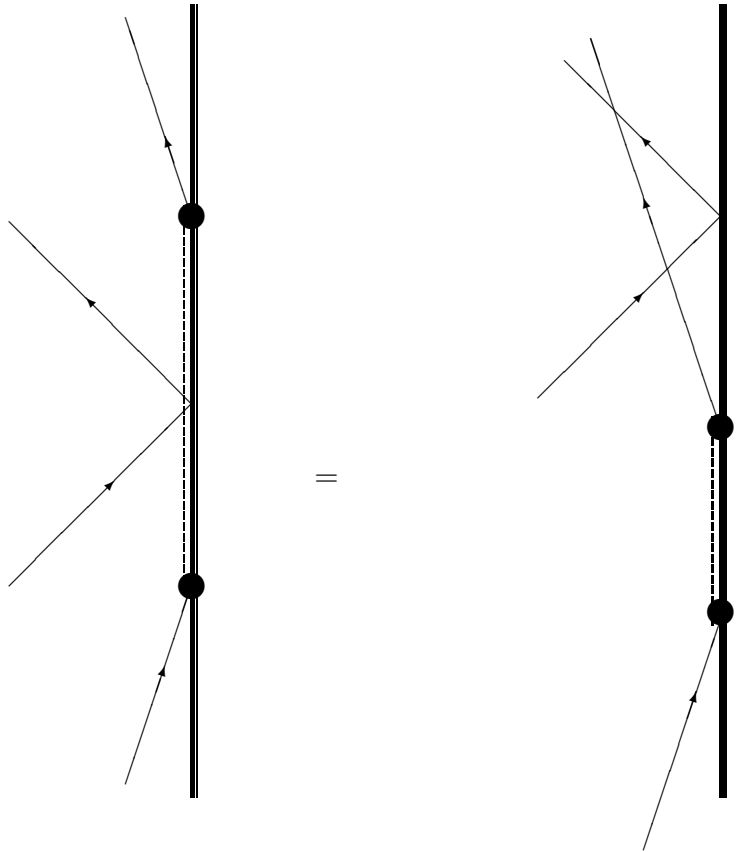
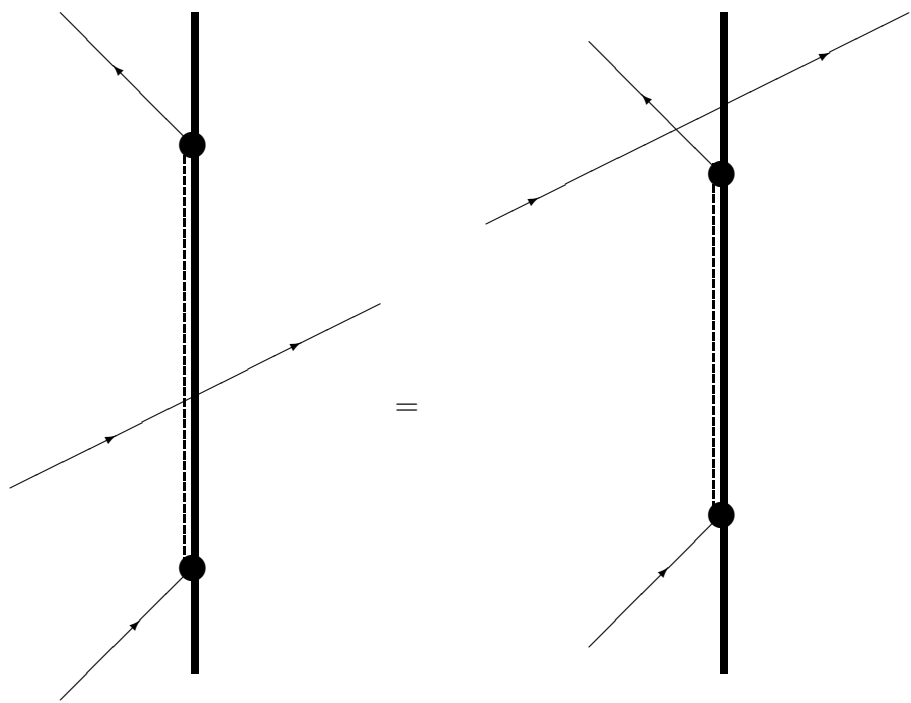
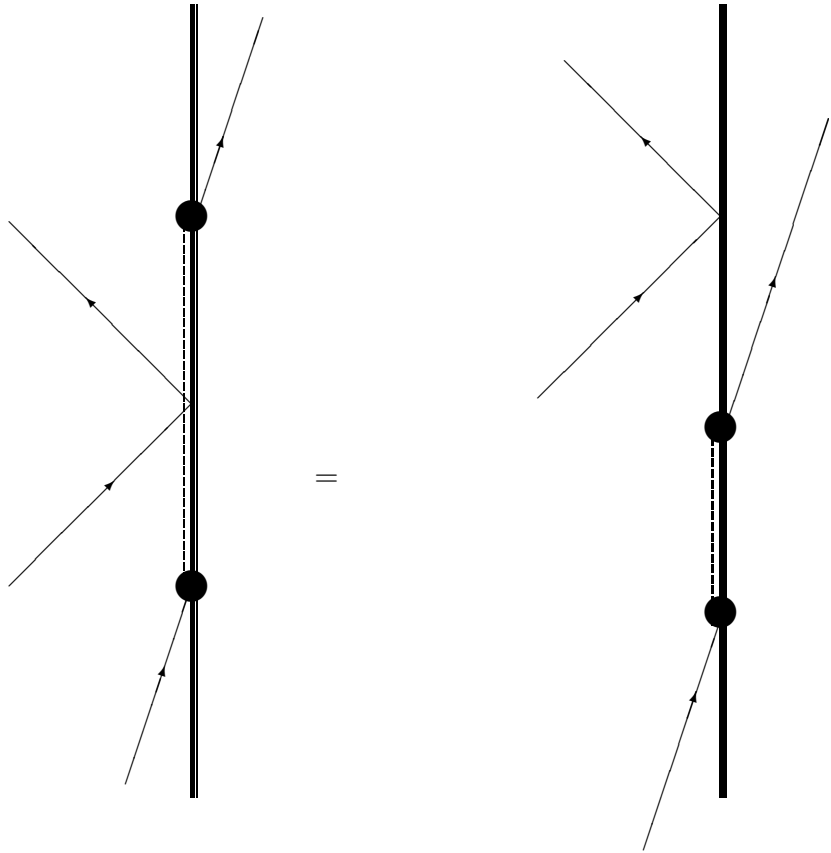
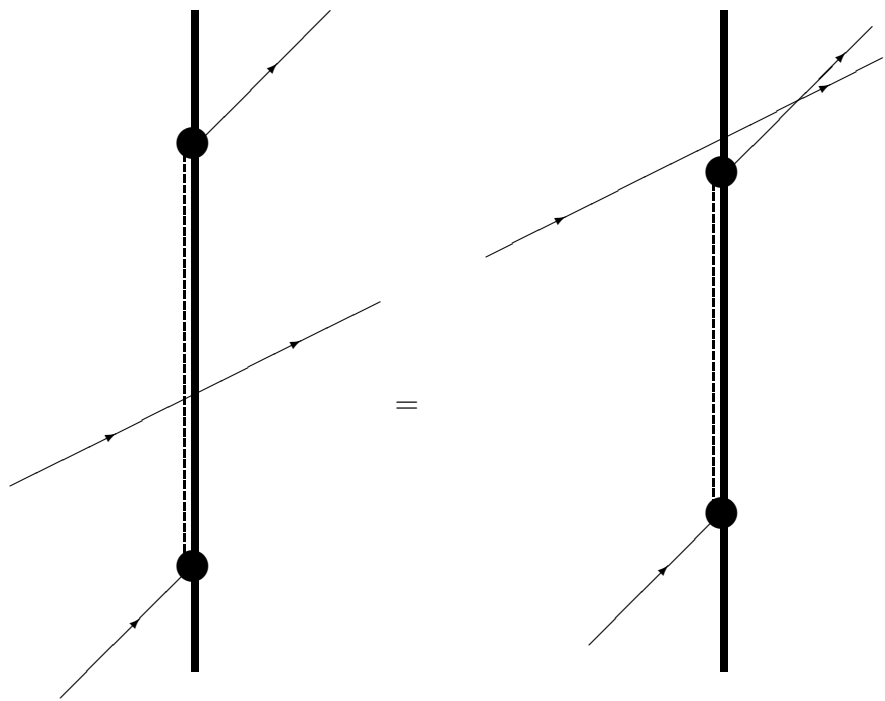
\item [Figure 4]. Bootstrap Equations of the defect bound state in the
        reflection ($a$ and $b$) and in the transmission channel ($c$ and $d$).
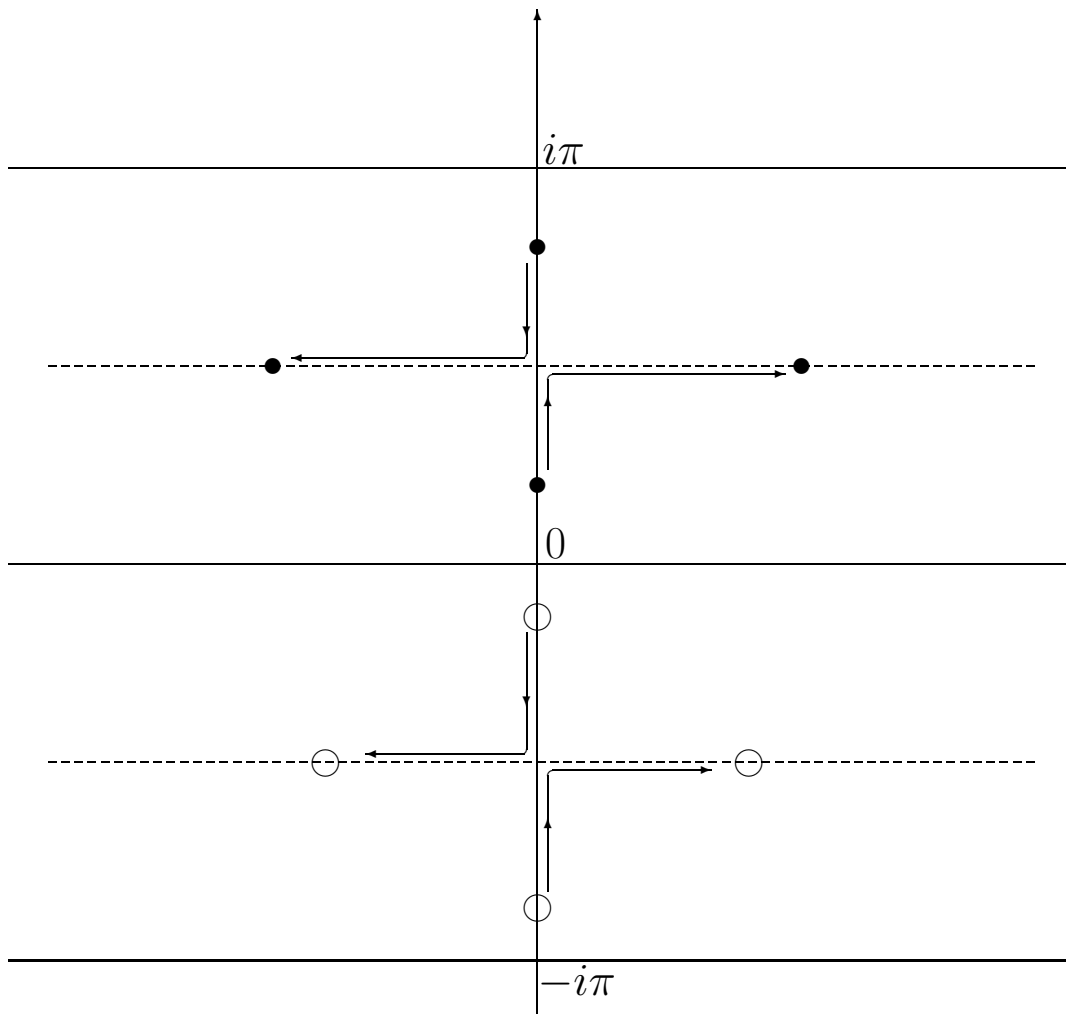
\item [Figure 5]. Pole structure of the bosonic amplitudes for
        positive values of the coupling constant (empty circles) and
        for negative ones (filled circles).
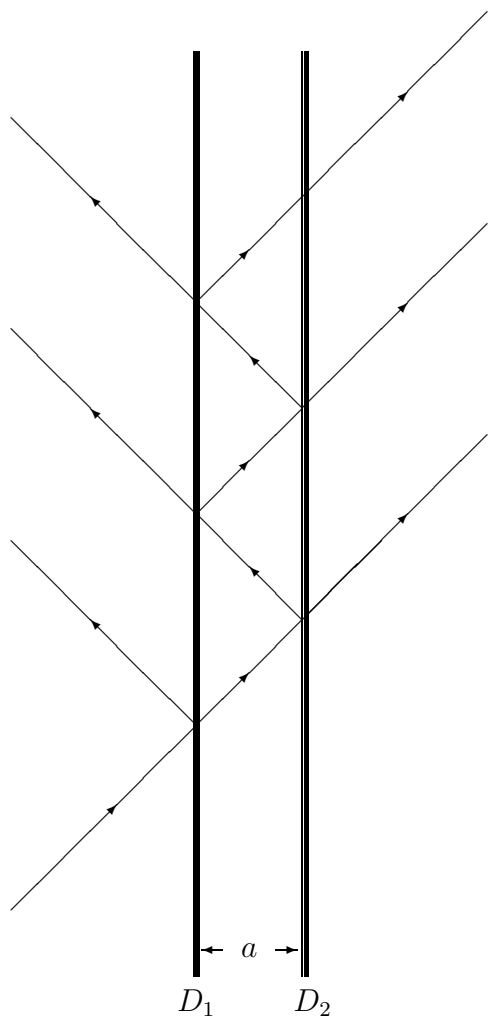
\item [Figure 6]. Scattering processes at the two defect lines.
\item [Figure 7]. Resonances in the transmission channel of the
        scattering on the two defect lines.
\item [Figure 8]. Band structure of the energy levels of the Majorana fermion
        with an infinite array of defects.
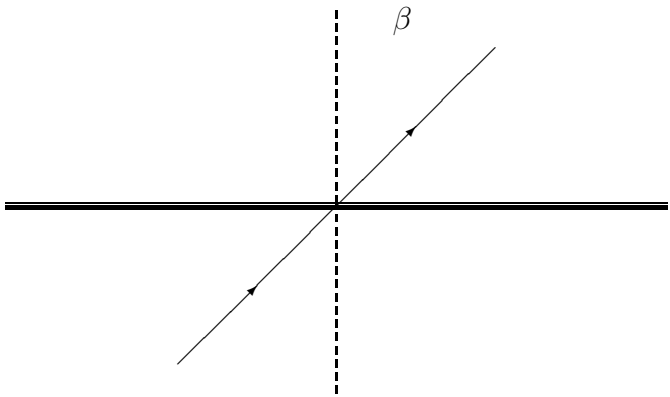
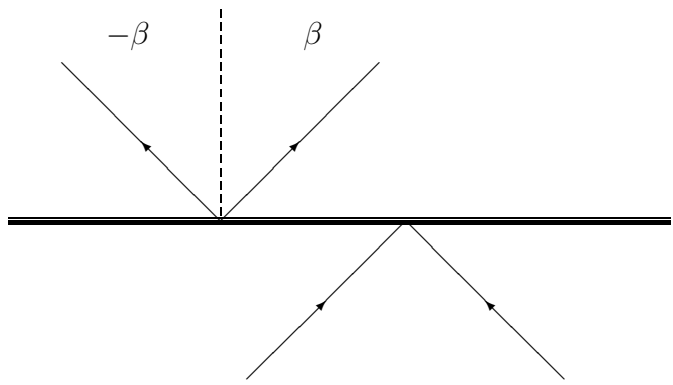
\item [Figure 9]. Defect line at $t=0$. Transmission channel ($a$) and
        processes of creation (annihilation) of a pair of particles ($b$).
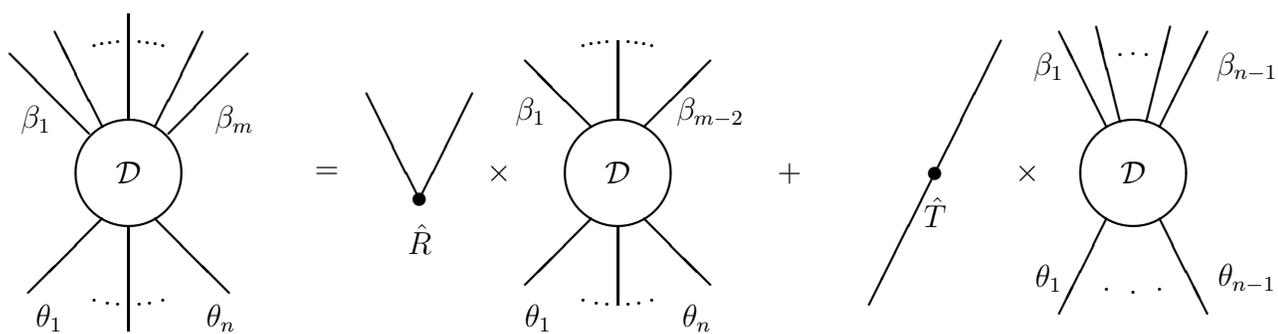
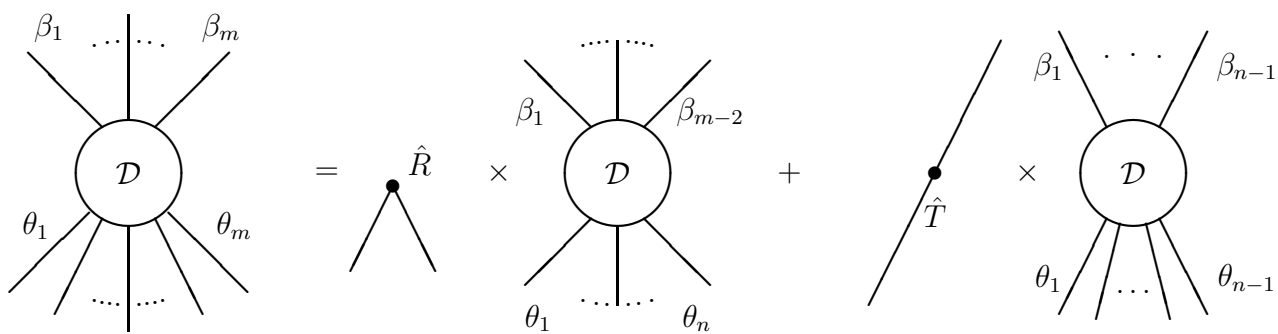
\item [Figure 10]. Recursive equations of the matrix elements of the operator
${\cal D}$.
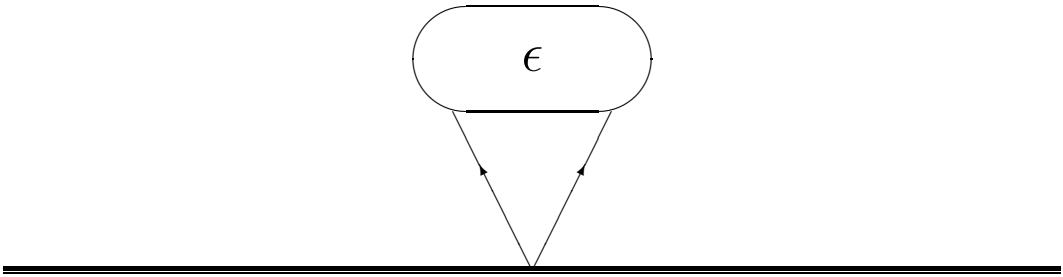
\item [Figure 11]. One point function of the energy operator of the Ising
        model in the presence of the defect line.
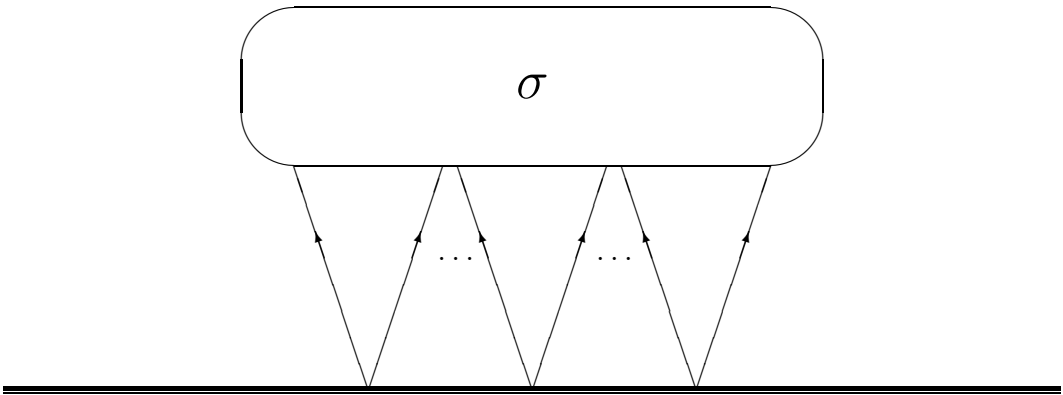
\item [Figure 12]. One point function of the magnetization operator of the
        Ising model in the presence of the defect line.

\end{description}

\newpage

\begin{picture}(400,500)
\put(100,200){\begin{picture}(200,200)
\put(99,10){\line(0,1){200}}
\put(101,10){\line(0,1){200}}
\put(100,10){\line(0,1){200}}
\put(100,0){\makebox(0,0){$ D$}}
\put(30,35){\vector(1,1){40}}
\put(70,75){\line(1,1){29}}
\put(70,53){\makebox(0,0){$A_i^{\dagger}(\beta)$}}
\put(99,106){\vector(-1,1){40}}
\put(59,146){\line(-1,1){29}}
\put(59,175){\makebox(0,0){$R_{ij}(\beta)$}}
\put(101,106){\vector(1,1){40}}
\put(141,146){\line(1,1){29}}
\put(141,175){\makebox(0,0){$T_{ij}(\beta)$}}
\end{picture}}
\end{picture}

\vspace{1mm}
\begin{center}
Figure 1
\end{center}

\newpage
\begin{picture}(300,500)
\put(50,100){\begin{picture}(300,400)
\put(99,0){\line(0,1){300}}
\put(101,0){\line(0,1){300}}
\put(100,0){\line(0,1){300}}
\put(30,30){\vector(1,1){40}}
\put(70,70){\line(1,1){29}}
\put(99,99){\vector(-1,1){40}}
\put(59,139){\line(-1,1){29}}
\put(40,0){\vector(1,3){15}}
\put(55,45){\line(1,3){5}}
\put(60,60){\line(1,3){39}}
\put(99,177){\vector(-1,3){20}}
\put(150,120){\makebox(0,0){$=$}}
\put(299,0){\line(0,1){300}}
\put(300,0){\line(0,1){300}}
\put(301,0){\line(0,1){300}}
\put(230,130){\vector(1,1){40}}
\put(270,170){\line(1,1){29}}
\put(299,199){\vector(-1,1){30}}
\put(269,229){\line(-1,1){29}}
\put(260,0){\vector(1,3){20}}
\put(280,60){\line(1,3){19}}
\put(299,117){\vector(-1,3){29}}
\put(270,204){\line(-1,3){20}}
\end{picture}}
\end{picture}
\vspace{1mm}
\begin{center}
\hspace{15mm} Figure 2.a
\end{center}

\newpage
\begin{picture}(400,500)
\put(0,100){\begin{picture}(400,400)
\put(99,0){\line(0,1){300}}
\put(101,0){\line(0,1){300}}
\put(100,0){\line(0,1){300}}
\put(30,30){\vector(1,1){40}}
\put(70,70){\line(1,1){29}}
\put(101,101){\vector(1,1){40}}
\put(141,141){\line(1,1){20}}
\put(40,0){\vector(1,3){15}}
\put(55,45){\line(1,3){5}}
\put(60,60){\line(1,3){39}}
\put(101,183){\vector(1,3){20}}
\put(121,243){\line(1,3){10}}
\put(160,120){\makebox(0,0){$=$}}
\put(299,0){\line(0,1){300}}
\put(300,0){\line(0,1){300}}
\put(301,0){\line(0,1){300}}
\put(230,130){\vector(1,1){40}}
\put(270,170){\line(1,1){29}}
\put(301,201){\vector(1,1){30}}
\put(331,231){\line(1,1){40}}
\put(260,0){\vector(1,3){20}}
\put(280,60){\line(1,3){19}}
\put(301,123){\vector(1,3){20}}
\put(321,183){\line(1,3){40}}
\end{picture}}
\end{picture}
\vspace{1mm}
\begin{center}
\hspace{5mm} Figure 2.b
\end{center}

\newpage
\begin{picture}(400,400)
\put(30,100){\begin{picture}(200,300)
\put(99,0){\line(0,1){200}}
\put(101,0){\line(0,1){200}}
\put(100,0){\line(0,1){200}}
\put(30,30){\vector(1,1){40}}
\put(70,70){\line(1,1){29}}
\put(99,99){\vector(-1,1){40}}
\put(59,139){\line(-1,1){29}}
\put(30,90){\vector(2,1){40}}
\put(70,110){\line(2,1){50}}
\put(120,135){\vector(2,1){20}}
\put(140,145){\line(2,1){30}}
\put(180,100){\makebox(0,0){$=$}}
\put(300,0){\line(0,1){200}}
\put(299,0){\line(0,1){200}}
\put(301,0){\line(0,1){200}}
\put(230,30){\vector(1,1){40}}
\put(270,70){\line(1,1){29}}
\put(299,99){\vector(-1,1){40}}
\put(259,139){\line(-1,1){29}}
\put(210,35){\vector(2,1){20}}
\put(230,45){\line(2,1){40}}
\put(270,65){\line(2,1){50}}
\put(320,90){\vector(2,1){20}}
\put(340,100){\line(2,1){30}}
\end{picture}}
\end{picture}

\vspace{1mm}
\begin{center}
\hspace{15mm} Figure 2.c
\end{center}

\newpage
\begin{picture}(300,500)
\put(30,100){\begin{picture}(300,400)
\put(99,0){\line(0,1){300}}
\put(101,0){\line(0,1){300}}
\put(100,0){\line(0,1){300}}
\put(30,30){\vector(1,1){40}}
\put(70,70){\line(1,1){29}}
\put(99,99){\vector(-1,1){40}}
\put(59,139){\line(-1,1){29}}
\put(40,0){\vector(1,3){15}}
\put(55,45){\line(1,3){5}}
\put(60,60){\line(1,3){39}}
\put(101,183){\vector(1,3){20}}
\put(121,243){\line(1,3){10}}
\put(160,120){\makebox(0,0){$=$}}
\put(299,0){\line(0,1){300}}
\put(300,0){\line(0,1){300}}
\put(301,0){\line(0,1){300}}
\put(230,130){\vector(1,1){40}}
\put(270,170){\line(1,1){29}}
\put(299,199){\vector(-1,1){30}}
\put(269,229){\line(-1,1){29}}
\put(260,0){\vector(1,3){20}}
\put(280,60){\line(1,3){19}}
\put(301,123){\vector(1,3){20}}
\put(321,183){\line(1,3){30}}
\end{picture}}
\end{picture}
\vspace{1mm}
\begin{center}
\hspace{5mm} Figure 2.d
\end{center}

\newpage
\vspace{3cm}
\begin{picture}(600,300)
\multiput(15,-10)(20,0){7}{\line(0,1){160}}
\multiput(0,10)(0,20){7}{\line(1,0){160}}
\put(74,-10){\line(0,1){160}}
\put(76,-10){\line(0,1){160}}
\put(45,80){\makebox(0,0){$J$}}
\put(65,40){\makebox(0,0){$\tilde J$}}

\multiput(325,-10)(20,0){7}{\line(0,1){160}}
\multiput(310,10)(0,20){7}{\line(1,0){160}}
\multiput(365,9)(0,20){7}{\line(1,0){20}}
\multiput(365,11)(0,20){7}{\line(1,0){20}}
\put(335,80){\makebox(0,0){$J$}}
\put(375,40){\makebox(0,0){$\tilde J$}}
\end{picture}

\vspace{15mm}
\begin{center}
\hspace{8mm}
Figure 3
\end{center}

\newpage
\begin{picture}(300,500)
\put(50,100){\begin{picture}(300,400)
\put(99,0){\line(0,1){300}}
\put(101,0){\line(0,1){300}}
\put(100,0){\line(0,1){300}}

\put(30,80){\vector(1,1){40}}
\put(70,120){\line(1,1){29}}
\put(99,149){\vector(-1,1){40}}
\put(59,189){\line(-1,1){29}}

\put(99,220){\vector(-1,3){10}}
\put(89,250){\line(-1,3){15}}
\put(99,220){\circle*{10}}

\put(74,5){\vector(1,3){15}}
\put(89,50){\line(1,3){10}}
\put(99,80){\circle*{10}}

\multiput(96,77)(0,4){37}{\line(0,1){3}}

\put(150,120){\makebox(0,0){$=$}}

\put(299,0){\line(0,1){300}}
\put(300,0){\line(0,1){300}}
\put(301,0){\line(0,1){300}}

\put(230,151){\vector(1,1){40}}
\put(270,191){\line(1,1){29}}
\put(299,220){\vector(-1,1){30}}
\put(269,250){\line(-1,1){29}}

\put(299,140){\vector(-1,3){29}}
\put(270,227){\line(-1,3){20}}
\put(299,140){\circle*{10}}

\put(270,-20){\vector(1,3){20}}
\put(290,40){\line(1,3){9}}
\put(299,70){\circle*{10}}

\multiput(296,65)(0,4){20}{\line(0,1){3}}
\end{picture}}
\end{picture}
\vspace{1mm}
\begin{center}
\hspace{15mm} Figure 4.a
\end{center}

\newpage
\begin{picture}(300,500)
\put(30,100){\begin{picture}(300,400)
\put(99,0){\line(0,1){250}}
\put(101,0){\line(0,1){250}}
\put(100,0){\line(0,1){250}}

\put(99,200){\vector(-1,1){30}}
\put(99,200){\circle*{10}}
\put(69,230){\line(-1,1){20}}
\put(49,-10){\vector(1,1){30}}
\put(79,20){\line(1,1){20}}
\put(99,40){\circle*{10}}
\multiput(96,37)(0,4){41}{\line(0,1){3}}

\put(30,70){\vector(2,1){40}}
\put(70,90){\line(2,1){50}}
\put(120,115){\vector(2,1){20}}
\put(140,125){\line(2,1){30}}

\put(180,100){\makebox(0,0){$=$}}

\put(300,0){\line(0,1){250}}
\put(299,0){\line(0,1){250}}
\put(301,0){\line(0,1){250}}

\put(299,189){\vector(-1,1){30}}
\put(299,189){\circle*{10}}
\put(269,219){\line(-1,1){20}}

\put(249,10){\vector(1,1){30}}
\put(279,40){\line(1,1){20}}
\put(299,60){\circle*{10}}
\multiput(296,58)(0,4){34}{\line(0,1){3}}

\put(210,170){\vector(2,1){20}}
\put(230,180){\line(2,1){40}}
\put(270,200){\line(2,1){50}}
\put(320,225){\vector(2,1){20}}
\put(340,235){\line(2,1){30}}
\end{picture}}
\end{picture}

\vspace{1mm}
\begin{center}
\hspace{15mm} Figure 4.b
\end{center}

\newpage
\begin{picture}(300,500)
\put(50,100){\begin{picture}(300,400)
\put(99,0){\line(0,1){300}}
\put(101,0){\line(0,1){300}}
\put(100,0){\line(0,1){300}}

\put(30,80){\vector(1,1){40}}
\put(70,120){\line(1,1){29}}
\put(99,149){\vector(-1,1){40}}
\put(59,189){\line(-1,1){29}}

\put(101,220){\vector(1,3){10}}
\put(99,220){\circle*{12}}
\put(111,250){\line(1,3){15}}

\put(74,5){\vector(1,3){15}}
\put(89,50){\line(1,3){10}}
\put(99,80){\circle*{12}}
\multiput(96,78)(0,4){35}{\line(0,1){3}}

\put(150,120){\makebox(0,0){$=$}}

\put(299,0){\line(0,1){300}}
\put(300,0){\line(0,1){300}}
\put(301,0){\line(0,1){300}}

\put(230,151){\vector(1,1){40}}
\put(270,191){\line(1,1){29}}
\put(299,220){\vector(-1,1){30}}
\put(269,250){\line(-1,1){29}}

\put(301,140){\vector(1,3){19}}
\put(299,140){\circle*{12}}
\put(320,197){\line(1,3){20}}

\put(270,-20){\vector(1,3){20}}
\put(290,40){\line(1,3){9}}
\put(299,70){\circle*{12}}
\multiput(296,65)(0,4){19}{\line(0,1){3}}
\end{picture}}
\end{picture}
\vspace{1mm}
\begin{center}
\hspace{15mm} Figure 4.c
\end{center}

\newpage
\begin{picture}(300,500)
\put(30,100){\begin{picture}(300,400)
\put(99,0){\line(0,1){250}}
\put(101,0){\line(0,1){250}}
\put(100,0){\line(0,1){250}}

\put(101,200){\vector(1,1){30}}
\put(99,200){\circle*{12}}
\put(131,230){\line(1,1){20}}

\put(49,-10){\vector(1,1){30}}
\put(79,20){\line(1,1){20}}
\put(99,40){\circle*{12}}
\multiput(96,37)(0,4){40}{\line(0,1){3}}

\put(30,70){\vector(2,1){40}}
\put(70,90){\line(2,1){50}}
\put(120,115){\vector(2,1){20}}
\put(140,125){\line(2,1){30}}

\put(180,100){\makebox(0,0){$=$}}

\put(300,0){\line(0,1){250}}
\put(299,0){\line(0,1){250}}
\put(301,0){\line(0,1){250}}

\put(301,189){\vector(1,1){45}}
\put(299,189){\circle*{12}}
\put(346,234){\line(1,1){10}}

\put(210,157){\vector(2,1){30}}
\put(240,172){\line(2,1){20}}
\put(260,182){\line(2,1){50}}
\put(310,207){\vector(2,1){30}}
\put(340,222){\line(2,1){20}}

\put(249,10){\vector(1,1){30}}
\put(279,40){\line(1,1){20}}
\put(299,60){\circle*{12}}
\multiput(296,58)(0,4){32}{\line(0,1){3}}
\end{picture}}
\end{picture}

\vspace{1mm}
\begin{center}
\hspace{15mm} Figure 4.d
\end{center}

\vspace{3cm}
\begin{picture}(500,500)
\put(20,0){\begin{picture}(500,500)
\put(0,20){\line(1,0){400}}
\put(0,170){\line(1,0){400}}
\put(0,320){\line(1,0){400}}
\put(207,178){\makebox(0,0){\mbox{\Large $0$}}}
\put(215,12){\makebox(0,0){\mbox{\Large $-i \pi$}}}
\put(210,327){\makebox(0,0){\mbox{\Large $i \pi$}}}
\put(200,0){\vector(0,1){380}}

\multiput(15,95)(5,0){75}{\line(1,0){3}}
\multiput(15,245)(5,0){75}{\line(1,0){3}}

\put(200,40){\circle{10}}
\put(200,150){\circle{10}}
\put(120,95){\circle{10}}
\put(280,95){\circle{10}}

\put(204,46){\vector(0,1){28}}
\put(204,74){\line(0,1){16}}
\put(206,90){\oval(4,4)[tl]}
\put(206,92){\line(1,0){32}}
\put(235,92){\vector(1,0){31}}

\put(196,144){\vector(0,-1){28}}
\put(196,116){\line(0,-1){16}}
\put(194,100){\oval(4,4)[br]}
\put(195,98){\line(-1,0){30}}
\put(165,98){\vector(-1,0){30}}

\put(200,200){\circle*{6}}
\put(200,290){\circle*{6}}
\put(100,245){\circle*{6}}
\put(300,245){\circle*{6}}

\put(204,206){\vector(0,1){28}}
\put(204,234){\line(0,1){6}}
\put(206,240){\oval(4,4)[tl]}
\put(206,242){\line(1,0){20}}
\put(226,242){\vector(1,0){68}}

\put(196,284){\vector(0,-1){28}}
\put(196,256){\line(0,-1){6}}
\put(194,250){\oval(4,4)[br]}
\put(195,248){\line(-1,0){20}}
\put(175,248){\vector(-1,0){68}}

\end{picture}}
\end{picture}

\vspace{3mm}
\begin{center}
\hspace{5mm} Figure 5
\end{center}

\newpage
\begin{picture}(500,750)
\put(100,180){\begin{picture}(400,450)
\put(99,50){\line(0,1){350}}
\put(101,50){\line(0,1){350}}
\put(100,50){\line(0,1){350}}
\put(30,75){\vector(1,1){40}}
\put(70,115){\line(1,1){30}}
\put(110,60){\vector(-1,0){8}}
\put(120,60){\makebox(0,0){$a$}}
\put(130,60){\vector(1,0){8}}
\put(100,145){\vector(-1,1){40}}
\put(60,185){\line(-1,1){30}}
\put(100,145){\vector(1,1){20}}
\put(120,165){\line(1,1){20}}
\put(140,185){\vector(1,1){40}}
\put(180,225){\line(1,1){30}}
\put(140,185){\vector(-1,1){20}}
\put(120,205){\line(-1,1){20}}
\put(100,225){\vector(-1,1){40}}
\put(100,225){\vector(1,1){20}}
\put(120,245){\line(1,1){20}}
\put(140,265){\vector(1,1){40}}
\put(180,305){\line(1,1){30}}
\put(140,265){\vector(-1,1){20}}
\put(120,285){\line(-1,1){20}}
\put(100,305){\vector(-1,1){40}}
\put(60,345){\line(-1,1){30}}
\put(100,305){\vector(1,1){20}}
\put(120,325){\line(1,1){20}}
\put(140,345){\vector(1,1){40}}
\put(180,385){\line(1,1){30}}
\put(60,265){\line(-1,1){30}}
\put(140,50){\line(0,1){350}}
\put(141,50){\line(0,1){350}}
\put(142,50){\line(0,1){350}}
\put(141,40){\makebox(0,0){$\,\,\, D_2$}}
\put(141,186){\line(1,1){29}}
\put(100,40){\makebox(0,0){$D_1$}}
\put(120,-10){\makebox(0,0){Figure 6}}
\end{picture}}
\end{picture}

\newpage

\begin{picture}(500,400)
\put(30,0){\begin{picture}(400,300)
\put(50,149){\line(1,0){250}}
\put(50,150){\line(1,0){250}}
\put(50,151){\line(1,0){250}}
\multiput(175,151)(0,5){15}{\line(0,1){3}}
\multiput(175,149)(0,-5){15}{\line(0,1){3}}
\put(200,220){\makebox(0,0){$\beta$}}
\put(175,150){\vector(1,1){30}}
\put(205,180){\line(1,1){30}}
\put(115,90){\vector(1,1){30}}
\put(145,120){\line(1,1){30}}
\end{picture}}
\end{picture}
\vspace{1mm}
\begin{center}
\hspace{5mm} Figure 9.a
\end{center}

\newpage
\begin{picture}(500,400)
\put(30,0){\begin{picture}(400,300)
\put(50,149){\line(1,0){250}}
\put(50,150){\line(1,0){250}}
\put(50,151){\line(1,0){250}}
\multiput(130,152)(0,5){16}{\line(0,1){3}}
\put(165,220){\makebox(0,0){$\beta$}}
\put(95,220){\makebox(0,0){$-\beta$}}
\put(130,150){\vector(1,1){30}}
\put(130,150){\vector(-1,1){30}}
\put(160,180){\line(1,1){30}}
\put(100,180){\line(-1,1){30}}
\put(140,90){\vector(1,1){30}}
\put(170,120){\line(1,1){30}}
\put(260,90){\vector(-1,1){30}}
\put(230,120){\line(-1,1){30}}
\end{picture}}
\end{picture}
\vspace{1mm}
\begin{center}
\hspace{5mm} Figure 9.b
\end{center}

\newpage

\begin{center}
\begin{picture}(450,200)(-45,0)
\thicklines
\put(-30,0){\circle{40}}
\put(-30,20){\line(0,1){40}}
\put(-30,-20){\line(0,-1){40}}
\put(-30,0){\makebox(0,0){${\cal D}$}}

\put(-45,15){\line(-1,1){30}}
\put(-15,15){\line(1,1){30}}
\put(-40,17.33){\line(-1,2){18}}
\put(-20,17.33){\line(1,2){18}}
\put(-40,-17.33){\line(-1,-1){28}}
\put(-20,-17.33){\line(1,-1){28}}
\put(-65,20){\makebox(0,0){$\beta_1$}}
\put(10,20){\makebox(0,0){$\beta_m$}}
\put(-60,-55){\makebox(0,0){$\theta_1$}}
\put(5,-55){\makebox(0,0){$\theta_n$}}
\put(-40,49){\makebox(0,0){$.$}}
\put(-37,49.5){\makebox(0,0){$.$}}
\put(-34,49.64){\makebox(0,0){$.$}}
\put(-43,48.30){\makebox(0,0){$.$}}
\put(-17,48.30){\makebox(0,0){$.$}}
\put(-20,49){\makebox(0,0){$.$}}
\put(-23,49.5){\makebox(0,0){$.$}}
\put(-26,49.64){\makebox(0,0){$.$}}
\put(-43,-48.30){\makebox(0,0){$.$}}
\put(-38,-49.35){\makebox(0,0){$.$}}
\put(-34,-49.83){\makebox(0,0){$.$}}
\put(-17,-48.30){\makebox(0,0){$.$}}
\put(-22,-49.35){\makebox(0,0){$.$}}
\put(-26,-49.83){\makebox(0,0){$.$}}

\put(45,0){\makebox(0,0){$=$}}

\put(80,-10){\circle*{5}}
\put(80,-25){\makebox(0,0){$\hat R$}}
\put(80,-10){\line(1,2){20}}
\put(80,-10){\line(-1,2){20}}
\put(110,0){\makebox(0,0){$\times$}}

\put(155,0){\circle{40}}
\put(155,0){\makebox(0,0){${\cal D}$}}
\put(165,-17.33){\line(1,-1){25}}
\put(155,20){\line(0,1){30}}
\put(155,-20){\line(0,-1){30}}
\put(145,17.33){\line(-1,1){25}}
\put(145,-17.33){\line(-1,-1){25}}
\put(165,17.33){\line(1,1){25}}

\put(122,23){\makebox(0,0){$\beta_1$}}
\put(190,23){\makebox(0,0){$\beta_{m-2}$}}
\put(125,-55){\makebox(0,0){$\theta_1$}}
\put(185,-55){\makebox(0,0){$\theta_n$}}
\put(145,49){\makebox(0,0){$.$}}
\put(148,49.5){\makebox(0,0){$.$}}
\put(151,49.64){\makebox(0,0){$.$}}
\put(142,48.30){\makebox(0,0){$.$}}
\put(168,48.30){\makebox(0,0){$.$}}
\put(165,49){\makebox(0,0){$.$}}
\put(162,49.5){\makebox(0,0){$.$}}
\put(159,49.64){\makebox(0,0){$.$}}
\put(142,-48.30){\makebox(0,0){$.$}}
\put(147,-49.35){\makebox(0,0){$.$}}
\put(151,-49.83){\makebox(0,0){$.$}}
\put(168,-48.30){\makebox(0,0){$.$}}
\put(163,-49.35){\makebox(0,0){$.$}}
\put(159,-49.83){\makebox(0,0){$.$}}

\put(220,0){\makebox(0,0){$+$}}

\put(250,-50){\line(1,2){50}}
\put(275,0){\circle*{5}}
\put(275,-15){\makebox(0,0){$\hat T$}}
\put(310,0){\makebox(0,0){$\times$}}

\put(350,0){\circle{40}}
\put(360,17.33){\line(1,2){18}}
\put(350,0){\makebox(0,0){${\cal D}$}}
\put(340,17.33){\line(-1,2){18}}
\put(345,19,36){\line(-1,4){9}}
\put(355,19.36){\line(1,4){9}}
\put(340,-17.33){\line(-1,-2){18}}
\put(360,-17.33){\line(1,-2){18}}
\put(318,40){\makebox(0,0){$\beta_1$}}
\put(393,40){\makebox(0,0){$\beta_{n-1}$}}
\put(318,-40){\makebox(0,0){$\theta_1$}}
\put(393,-40){\makebox(0,0){$\theta_{n-1}$}}
\put(350,45){\makebox(0,0){$.$}}
\put(345,44.7){\makebox(0,0){$.$}}
\put(355,44.7){\makebox(0,0){$.$}}
\put(350,-45){\makebox(0,0){$.$}}
\put(340,-43.9){\makebox(0,0){$.$}}
\put(360,-43.9){\makebox(0,0){$.$}}

\end{picture}
\end{center}
\vspace{35mm}
\begin{center}
{\bf Figure 10.a}
\end{center}

\newpage

\begin{center}
\begin{picture}(450,200)(-45,0)
\thicklines
\put(-30,0){\circle{40}}
\put(-30,20){\line(0,1){40}}
\put(-30,-20){\line(0,-1){40}}
\put(-30,0){\makebox(0,0){${\cal D}$}}
\put(-45,-15){\line(-1,-1){30}}
\put(-15,-15){\line(1,-1){30}}
\put(-40,-17.33){\line(-1,-2){18}}
\put(-20,-17.33){\line(1,-2){18}}
\put(-40,+17.33){\line(-1,1){28}}
\put(-20,+17.33){\line(1,1){28}}
\put(-65,-20){\makebox(0,0){$\theta_1$}}
\put(10,-20){\makebox(0,0){$\theta_m$}}
\put(-60,55){\makebox(0,0){$\beta_1$}}
\put(5,55){\makebox(0,0){$\beta_m$}}
\put(-40,-49){\makebox(0,0){$.$}}
\put(-37,-49.5){\makebox(0,0){$.$}}
\put(-34,-49.64){\makebox(0,0){$.$}}
\put(-43,-48.30){\makebox(0,0){$.$}}
\put(-17,-48.30){\makebox(0,0){$.$}}
\put(-20,-49){\makebox(0,0){$.$}}
\put(-23,-49.5){\makebox(0,0){$.$}}
\put(-26,-49.64){\makebox(0,0){$.$}}
\put(-43,48.30){\makebox(0,0){$.$}}
\put(-38,49.35){\makebox(0,0){$.$}}
\put(-34,49.83){\makebox(0,0){$.$}}
\put(-17,48.30){\makebox(0,0){$.$}}
\put(-22,49.35){\makebox(0,0){$.$}}
\put(-26,49.83){\makebox(0,0){$.$}}

\put(45,0){\makebox(0,0){$=$}}

\put(70,-5){\circle*{5}}
\put(80,5){\makebox(0,0){$\hat R$}}
\put(70,-5){\line(1,-2){16}}
\put(70,-5){\line(-1,-2){16}}
\put(110,0){\makebox(0,0){$\times$}}

\put(155,0){\circle{40}}
\put(155,0){\makebox(0,0){${\cal D}$}}
\put(165,-17.33){\line(1,-1){25}}
\put(155,20){\line(0,1){30}}
\put(155,-20){\line(0,-1){30}}
\put(145,17.33){\line(-1,1){25}}
\put(145,-17.33){\line(-1,-1){25}}
\put(165,17.33){\line(1,1){25}}

\put(122,23){\makebox(0,0){$\beta_1$}}
\put(190,23){\makebox(0,0){$\beta_{m-2}$}}
\put(125,-55){\makebox(0,0){$\theta_1$}}
\put(185,-55){\makebox(0,0){$\theta_n$}}
\put(145,49){\makebox(0,0){$.$}}
\put(148,49.5){\makebox(0,0){$.$}}
\put(151,49.64){\makebox(0,0){$.$}}
\put(142,48.30){\makebox(0,0){$.$}}
\put(168,48.30){\makebox(0,0){$.$}}
\put(165,49){\makebox(0,0){$.$}}
\put(162,49.5){\makebox(0,0){$.$}}
\put(159,49.64){\makebox(0,0){$.$}}
\put(142,-48.30){\makebox(0,0){$.$}}
\put(147,-49.35){\makebox(0,0){$.$}}
\put(151,-49.83){\makebox(0,0){$.$}}
\put(168,-48.30){\makebox(0,0){$.$}}
\put(163,-49.35){\makebox(0,0){$.$}}
\put(159,-49.83){\makebox(0,0){$.$}}

\put(220,0){\makebox(0,0){$+$}}

\put(250,-50){\line(1,2){50}}
\put(275,0){\circle*{5}}
\put(275,-15){\makebox(0,0){$\hat T$}}
\put(310,0){\makebox(0,0){$\times$}}

\put(350,0){\circle{40}}
\put(360,-17.33){\line(1,-2){18}}
\put(340,17.33){\line(-1,2){18}}
\put(350,0){\makebox(0,0){${\cal D}$}}
\put(340,-17.33){\line(-1,-2){18}}
\put(345,-19,36){\line(-1,-4){9}}
\put(355,-19.36){\line(1,-4){9}}
\put(360,17.33){\line(1,2){18}}

\put(318,-40){\makebox(0,0){$\theta_1$}}
\put(393,-40){\makebox(0,0){$\theta_{n-1}$}}
\put(318,40){\makebox(0,0){$\beta_1$}}
\put(393,40){\makebox(0,0){$\beta_{n-1}$}}

\put(350,-45){\makebox(0,0){$.$}}
\put(345,-44.7){\makebox(0,0){$.$}}
\put(355,-44.7){\makebox(0,0){$.$}}
\put(350,45){\makebox(0,0){$.$}}
\put(340,43.9){\makebox(0,0){$.$}}
\put(360,43.9){\makebox(0,0){$.$}}

\end{picture}
\end{center}
\vspace{35mm}
\begin{center}
{\bf Figure 10.b}
\end{center}

\newpage

\begin{picture}(500,500)
\put(20,0){\begin{picture}(500,500)

\put(0,170){\line(1,0){400}}
\put(0,169){\line(1,0){400}}
\put(0,171){\line(1,0){400}}

\put(200,250){\oval(90,40)}
\put(200,250){\makebox(0,0){\mbox{\LARGE $\epsilon$}}}

\put(200,170){\vector(1,2){20}}
\put(220,210){\line(1,2){10}}
\put(200,170){\vector(-1,2){20}}
\put(180,210){\line(-1,2){10}}

\end{picture}}
\end{picture}

\vspace{3mm}
\begin{center}
\hspace{5mm} Figure 11
\end{center}

\newpage

\begin{picture}(500,500)
\put(20,0){\begin{picture}(500,500)

\put(0,170){\line(1,0){400}}
\put(0,169){\line(1,0){400}}
\put(0,171){\line(1,0){400}}

\put(200,285){\oval(220,60)}
\put(200,285){\makebox(0,0){\mbox{\LARGE $\sigma$}}}

\put(138,170){\vector(1,3){20}}
\put(158,230){\line(1,3){8.33}}
\put(138,170){\vector(-1,3){20}}
\put(118,230){\line(-1,3){8.33}}

\put(200,170){\vector(1,3){20}}
\put(220,230){\line(1,3){8.33}}
\put(200,170){\vector(-1,3){20}}
\put(180,230){\line(-1,3){8.33}}

\put(171,220){\makebox(0,0){$\ldots$}}
\put(231,220){\makebox(0,0){$\ldots$}}

\put(262,170){\vector(1,3){20}}
\put(282,230){\line(1,3){8.33}}
\put(262,170){\vector(-1,3){20}}
\put(242,230){\line(-1,3){8.33}}

\end{picture}}
\end{picture}

\vspace{3mm}
\begin{center}
\hspace{5mm} Figure 12
\end{center}

\end{document}